\documentclass[%
 aip,
 amsmath,amssymb,
 reprint,%
]{revtex4-1}

\usepackage{graphicx}
\usepackage{bm}
\usepackage{booktabs}
\usepackage{float}
\usepackage{xcolor}
\usepackage{hyperref}
\usepackage{cleveref}

\crefname{equation}{Eq.}{Eqs.}      
\Crefname{equation}{Equation}{Equations}  

\crefname{figure}{Fig.}{Figs.}
\Crefname{figure}{Figure}{Figures}

\crefname{table}{Table}{Tables}
\Crefname{table}{Table}{Tables}

\crefname{section}{Sec.}{Secs.}
\Crefname{section}{Section}{Sections}

\hypersetup{
    colorlinks=true,
    linkcolor=blue!60!black,
    citecolor=blue!60!black,
    urlcolor=blue!60!black
}

\begin{document}

\preprint{AIP/123-QED}

\title[A Graph Neural Network Framework for Characterizing Rainfall Variability Regimes
across India]{A Graph Neural Network Framework for\\
Characterizing Rainfall Variability Regimes
across India}

\author{Pradyumnan Raghuveeran}
\affiliation{Department of Aerospace Engineering, Indian Institute of Technology Madras}
\affiliation{Centre of Excellence for studying Critical Transitions in Complex Systems}

\author{Gaurav Chopra}
\affiliation{Department of Applied Mechanics, Indian Institute of Technology Delhi}

\author{Ajay Bankar}
\affiliation{Department of Aerospace Engineering, Indian Institute of Technology Madras}
\affiliation{Centre of Excellence for studying Critical Transitions in Complex Systems}

\author{R. I. Sujith}
 \email{sujith@iitm.ac.in}
\affiliation{Department of Aerospace Engineering, Indian Institute of Technology Madras}
\affiliation{Centre of Excellence for studying Critical Transitions in Complex Systems}

\date{\today}

\begin{abstract}
The Indian Summer Monsoon exhibits pronounced spatial heterogeneity, with rainfall
amount and onset dates differing sharply across the
subcontinent. While extensive prior work has characterized this heterogeneity using linear
statistical decompositions or forecasting-oriented deep learning models that predict
future rainfall values, comparatively little attention has been paid to
characterizing the repeatability of a location's rainfall pattern across
years, that is, how consistently the same seasonal rainfall trajectory recurs from
year to year, independent of forecasting its exact future value. We
propose a graph-based machine learning framework that classifies individual grid
points across India according to the inter-annual consistency of their daily rainfall
patterns. At each of 29,026 grid points from the GSMaP ISRO dataset (2001--2022,
excluding 2012), a graph is constructed in which nodes represent individual years of
daily rainfall, plus a climatological node representing the long-term mean, with edges
defined by a cosine-similarity between year-pairs. A Graph Convolutional
Network is trained on a sample of these graphs to classify locations as
either consistent (exhibiting a repeatable seasonal rainfall trajectory across
years) or erratic (lacking such repeatability), achieving 96.8\% validation
accuracy. Applied across the full Indian landmass, the classification identifies the
Western Ghats, Northeast India, and parts of central India as consistent regions,
consistent with known monsoon climatology. We validate the classification through statistical hypothesis testing against
independent climatological metrics, graph-structural analysis, threshold-sensitivity
analysis, and temporal-stability analysis across two independent periods
(2001--2011 and 2013--2022). Together, these confirm that the classification
reflects genuine physical structure, with 93.6\% agreement between the two periods. The results further reveal a previously unreported coupling between intra-annual
rainfall amplitude and inter-annual pattern consistency, an emergent, spatially
coherent structure not imposed by the classification procedure, indicating that
climatologically rainfall-rich regions are also the most temporally repeatable.
\end{abstract}

\maketitle

\begin{quotation}
\textbf{Rainfall over India, driven by the summer monsoon, is famously irregular. Some
places receive rain that follows almost the same pattern every year, while others
see highly variable rainfall behavior from one year to the next. Distinguishing
these two kinds of places matters for farmers, water-resource planners, and anyone trying to
prepare for droughts or floods, yet most existing approaches summarize
rainfall using statistical measures,
without asking a more basic question: how repeatable is a location's year-to-year
rainfall pattern in the first place? Here, we treat this as a problem of converting
a temporal pattern into a network. At each of nearly thirty thousand locations
across India, we compare each year's daily rainfall pattern to every other year's,
building a small network in which years that behaved similarly are linked together.
A location whose years are all mutually similar forms a densely connected network
and is called consistent. A location whose years differ substantially from one
another forms a sparse network and is called erratic. We then train a graph-based
machine learning model to recognize these network patterns and use it to classify
every location across the country. Strikingly, the resulting map is not scattered or
random. Rather, it organizes itself into large, geographically coherent regions, even though
the model was never given any information about geography. We further find that regions with
large swings in rainfall between the wet and dry seasons, typically the wettest
parts of the country, are also the most repeatable from year to year, while drier
regions tend to be more erratic. This reveals an unexpected coupling between two seemingly distinct aspects of rainfall variability. More broadly, this
work illustrates how tools from network science and graph learning can transform
noisy, high-dimensional time series into interpretable structures, offering a
template that extends beyond rainfall to other systems where the central question is
not what happens next, but whether what happens tends to repeat.}
\end{quotation}

\section{Introduction}
\label{sec:intro}

The Indian Summer Monsoon (ISM) is the dominant climatic phenomenon governing the socioeconomic fabric of South Asia, supplying roughly 70--90\% of India's annual rainfall within the June--September season \citep{yadav2025recent, ghosh2016indian, verma2022regional, hrudya2021review, sahu2026climate, thota2024spatial}. Its behavior cascades directly into agricultural productivity, water resource management, hydropower generation, and macroeconomic stability across the region \citep{ghosh2016indian, sahastrabuddhe2023indian, kathayat2022protracted, rajbanshi2021variability, turner2012climate, wang2006asian}. Indian agriculture remains substantially rain-fed, because of which deviations in seasonal rainfall totals or timing translate directly into reduced crop yields, rural income loss, and measurable impacts on GDP, with the agricultural sector alone contributing an estimated 13.7--20\% of India's GDP and employing a majority of its rural workforce \citep{rajbanshi2021variability, kathayat2022protracted, turner2012climate}. Beyond agriculture, monsoon variability has direct bearing on drought and flood risk, reservoir operation, and the socioeconomic well-being of over a billion people across the subcontinent \citep{mishra2022framework, verma2022regional, kapa2025variability, yadav2022monsoon}. Distinguishing locations with stable, repeatable year-to-year rainfall behavior from those with erratic, unpredictable patterns is itself of direct practical value: it informs where seasonal forecasts are likely to be reliable, where water storage and irrigation infrastructure should be prioritized as a hedge against unpredictability, how agricultural and crop-insurance risk should be assessed beyond simple rainfall totals, and how drought and flood early-warning thresholds should be calibrated regionally rather than applied uniformly across a climatically heterogeneous country.

Critically, the ISM is not spatially uniform. Seasonal rainfall amount, onset and withdrawal timing, and inter-annual variability differ sharply across the subcontinent: the West Coast (Western Ghats) and Northeast India typically receive the highest seasonal totals, while the northwestern arid zone and parts of the Deccan Plateau receive substantially less rainfall and exhibit markedly higher year-to-year variability \citep{hrudya2021review, ghosh2016indian, sahastrabuddhe2023indian, thota2024spatial, dalai2025deciphering, hegde2025spatio}. Regional trend analyses further reveal contrasting, and at times opposing, long-term rainfall trends across sub-regions, with parts of western India and the Western Ghats showing increasing rainfall while the Gangetic Plain and Northeast India show declining trends \citep{yadav2025recent, kathayat2022protracted, dalai2025deciphering}. Fine-resolution gridded analyses confirm that this spatial heterogeneity extends down to sub-degree scales, with the magnitude of inter-annual variability and the presence of significant temporal trends varying considerably even within nominally homogeneous monsoon regions \citep{duncan2013analysing, saini2022unraveling, mishra2022framework, saipriya2026space}. This heterogeneity means that national or basin-averaged rainfall indices, long the default target of monsoon prediction and monitoring efforts, can obscure regionally distinct behavior that is of direct relevance to water resource planners and agricultural policymakers operating at sub-national scales \citep{mishra2022framework, obata2024earth}. Characterizing where and how rainfall variability differs across India, in a manner that is both data-driven and physically interpretable, therefore remains an open and practically consequential problem in monsoon science.

A substantial body of work has characterized ISM inter-annual variability using classical statistical and dynamical approaches. Empirical Orthogonal Function (EOF) and extended EOF analyses have been widely used to decompose the ISM rainfall variability into dominant modes, revealing that different sub-regions of India often exhibit opposing, out-of-phase rainfall anomalies, with the El Ni\~no--Southern Oscillation (ENSO), the Indian Ocean Dipole (IOD), and the monsoon trough identified as primary modulators \citep{hrudya2021review, kathayat2022protracted, dalai2025deciphering, yadav2022monsoon, suhas2013indian}. Related studies using gridded and reanalysis rainfall products have documented contrasting long-term trends in the spatial distribution of both mean rainfall and rainfall extremes across India, distinguishing regions of increasing versus decreasing rainfall intensity and linking these trends to shifts in drought and flood risk \citep{yadav2025recent, ghosh2016indian, verma2022regional, mishra2022framework}. Other approaches have focused on identifying active and break monsoon spells from area-averaged rainfall indices \citep{suhas2013indian, saha2020cnn, hussain2021survey}, and on regionalizing India into broadly homogeneous monsoon zones using clustering of station or gridded rainfall records \citep{mishra2022framework, thota2024spatial, saini2022unraveling}. These approaches have been foundational in establishing the large-scale spatial structure of ISM variability, but they largely rely on linear decompositions, fixed regionalization schemes, or hand-designed indices, which may not fully capture more complex, nonlinear structure in year-to-year rainfall pattern similarity at fine spatial resolution.

More recently, machine learning (ML) and deep learning (DL) methods have been applied extensively to monsoon-related prediction and pattern-detection tasks. Recurrent architectures, particularly Long Short-Term Memory (LSTM) networks and stacked autoencoders, have been used to detect monsoon active and break spells and to identify skillful climatic predictors of seasonal rainfall \citep{saha2017deep, saha2021prediction, saha2016predictor, hussain2021survey}. Convolutional neural network (CNN)-based and hybrid CNN--LSTM architectures have demonstrated skill in forecasting all-India and regional seasonal rainfall, including at extended lead times of up to two years, by leveraging spatial patterns in sea-surface temperature, sub-surface ocean heat content, and other atmospheric predictor fields \citep{sharma2026improving, patil2025enhancing, sharma2022mechanism, kumar2022deep, kumar2026rainfall}. Transformer-based and ConvLSTM-based models have further been used for intraseasonal oscillation forecasting and short-range precipitation prediction integrating pre-monsoon and satellite-derived features \citep{anirudh2025skillful, bisht2026deep, kumar2022deep}. Ensemble and comparative studies have benchmarked classical ML, statistical, and deep learning models against one another for both all-India and regional monsoon forecasting, consistently finding that deep learning architectures outperform traditional regression-based and tree-based methods \citep{narang2024artificial, goyal2023conception, kumar2025forecasting, dash2026performance, bajpai2023deep, dash2024integrating, talan2026machine}. A common thread across this body of work is that rainfall is typically represented either as a single univariate seasonal or regional index, or as a gridded spatial field evolving in time, with the learning objective framed as forecasting future rainfall values. Comparatively little attention has been paid to using ML to characterize the structural similarity of rainfall patterns across years at the level of individual locations, i.e., treating each year's rainfall trajectory as an object to be compared and classified, rather than a quantity to be predicted.

Graph Neural Networks (GNNs) have emerged as a natural framework for modeling relational and non-Euclidean structure in climate and weather data \citep{li2023graph, lam2023learning, sun2025utility}. Unlike standard feed-forward or convolutional architectures, which treat samples independently or assume a fixed grid topology, GNNs explicitly represent entities as nodes and their pairwise relationships as edges, allowing information to propagate across the graph and enabling representations that jointly capture local node attributes and global connectivity structure \citep{li2023graph, sun2025utility, keisler2022forecasting}. This flexibility has motivated a rapidly growing number of applications of GNNs to climate and Earth-system problems. At the global scale, GNN-based emulators such as GraphCast have achieved skillful medium-range weather forecasting by operating on multi-resolution mesh graphs \citep{lam2023learning, keisler2022forecasting}. At regional and station scales, spatiotemporal GNNs have been used for precipitation forecasting and correction over irregular gauge and station networks \citep{zhang2023st, yousaf2025spatio, zheng2026mesh, li2026navigating, devkota2025spatio, xu2024dgformer, ma2023histgnn, wang2026spatiotemporal}, for improving numerical weather prediction of rainfall through structured feature propagation \citep{peng2023structured, chen2024coupling}, and for statistical and dynamical downscaling of precipitation and temperature to convection-permitting resolutions \citep{yan2025convolutional, blasone2025graph, blasone2026graph, coppola2025graph, blasone2024deep}. GNNs have also been applied in data-scarce settings to impute and forecast missing atmospheric variables \citep{bhandari2024recent}. Of particular relevance to the present work, Romanova \citep{romanova2024gnn} constructed graphs of year-wise daily temperature vectors at the city level, with edges defined by cosine similarity between years, and used a Graph Convolutional Network (GCN) \citep{kipf2016semi} to detect anomalous, climate-change-driven shifts in temperature patterns. This construction, representing each year as a node characterized by its full daily time series and connected to other years via a similarity metric, offers a template for encoding inter-annual pattern consistency directly into a graph structure, rather than relying on hand-designed summary statistics or fixed-window indices.

Despite these advances, three gaps remain. First, classical statistical characterizations of ISM spatial heterogeneity, whether EOF-based, trend-based, or clustering-based, are largely linear and rely on pre-specified indices or fixed regions, limiting their ability to capture fine-grained, grid-point-level distinctions in inter-annual rainfall consistency across the full Indian landmass. Second, existing ML/DL applications to Indian monsoon rainfall are overwhelmingly forecasting-oriented, targeting future rainfall values rather than characterizing the degree to which a location's year-to-year rainfall evolution is self-similar or erratic. This is a distinct and complementary question: knowing how consistent a location's rainfall pattern is, independent of predicting its exact future value, has direct relevance to water resource planning, agricultural risk assessment, and drought/flood early-warning system design. Third, while GNNs have been applied extensively to weather forecasting, precipitation downscaling, and station-network modeling, Romanova \citep{romanova2024gnn} demonstrated that year-wise similarity graphs can detect anomalous temporal patterns for temperature at city scale. This graph-based paradigm has not, to our knowledge, been applied to daily rainfall at fine spatial resolution across a large, climatically heterogeneous domain such as India, nor has it been accompanied by a multi-faceted validation framework establishing that the resulting classification reflects genuine physical structure rather than an artifact of graph construction.

To address this gap, we reframe the problem of characterizing rainfall consistency
not as a classification task performed directly on rainfall values, but as one of
converting temporal variability into network topology. Daily rainfall at each
location is first organized into year-wise trajectories. The pairwise similarity of
these trajectories defines a network whose topology, its connectivity, density, and
structure, encodes how consistently that location's rainfall repeats from year to
year. A GNN is then used to learn from this topology directly, rather than from the
raw rainfall values themselves. We propose a graph-based machine learning framework
that classifies individual grid points across India according to the inter-annual
consistency of their daily rainfall patterns. For each location, we construct a graph in which nodes represent individual years, each associated with a 365-dimensional feature vector of log-transformed daily rainfall, together with an additional climatological node representing the long-term mean rainfall cycle. Edges are established between year-pairs whose rainfall vectors exceed a global cosine-similarity threshold, following the similarity-graph paradigm introduced for temperature time series by Romanova \citep{romanova2024gnn}, but adapted here to the strongly right-skewed statistics of daily rainfall via a logarithmic transform, which bounds cosine similarity in $[0,1]$ and makes a global percentile threshold directly interpretable as a minimum pattern-similarity criterion. A Graph Convolutional Network \citep{kipf2016semi}, implemented using PyTorch Geometric \citep{fey2019fast}, is trained on a stratified sample of these per-location graphs to classify locations as consistent or erratic, and the trained model is then applied across all 29,026 grid points over the Indian landmass to produce a spatially continuous classification. Unlike prior forecasting-oriented applications of LSTMs and CNNs to ISM rainfall \citep{sharma2026improving, saha2020cnn, kumar2022deep}, our objective is characterization rather than prediction: we ask which locations exhibit stable, repeatable year-to-year rainfall evolution, and which do not. To establish that this classification reflects genuine climatological structure rather than a mathematical artifact of graph construction, we introduce a multi-faceted validation framework combining statistical hypothesis testing against independent physical metrics (intra-annual variance, coefficient of variation, rainy-day counts), graph-structural analysis (degree, clustering coefficient, density), threshold-sensitivity analysis, and temporal-stability analysis across two independent periods (2001--2011 and 2013--2022).

Beyond the classification itself, our results reveal a previously unreported coupling
between intra-annual rainfall amplitude and inter-annual pattern consistency,
suggesting that these two, ostensibly distinct, notions of rainfall variability are
physically linked rather than independent.

The remainder of this paper is structured as follows. \Cref{sec:datasets and methodology} describes the GSMaP ISRO dataset and the logarithmic preprocessing applied to the rainfall data, details the graph construction procedure, labeling strategy, and GNN architecture. \Cref{sec:results} presents the classification results across the Indian landmass and examines the relationship between intra- and inter-annual variability. \Cref{sec:validation} provides statistical, structural, threshold, and temporal validation of the classification. \Cref{sec:discussion} discusses the physical interpretation of the results and methodological caveats, and \cref{sec:conclusion} summarizes the key findings.

\section{Datasets and Methodology}
\label{sec:datasets and methodology}

\subsection{Datasets}
\label{subsec:datasets}

\begin{figure}[htb]
    \centering
    \includegraphics[width=\linewidth]{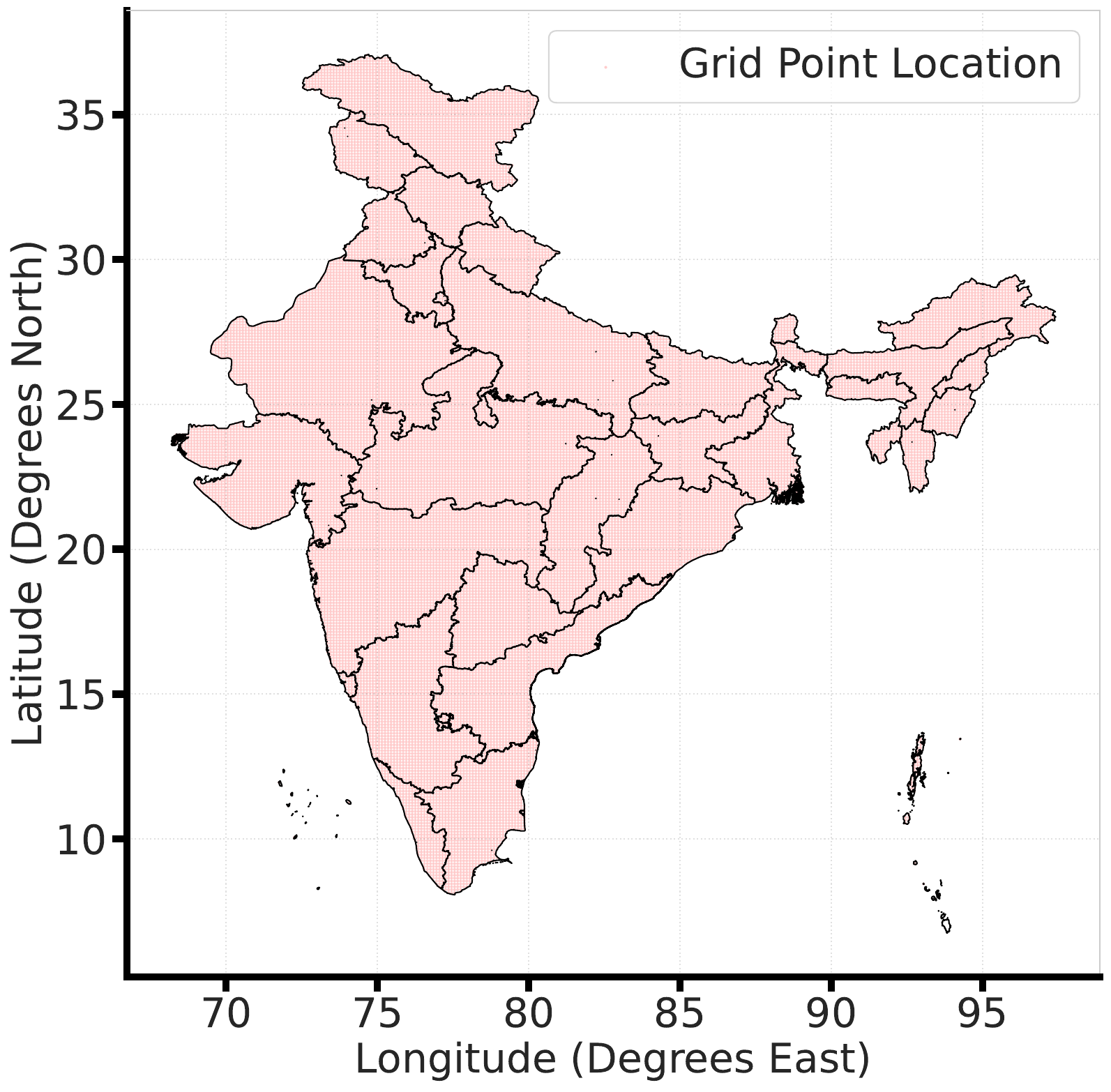}
    \caption{Spatial distribution of the 29,026 GSMaP ISRO grid points used in this
study, restricted to the Indian landmass. Each point marks the location of a
$0.1^\circ$ ($\sim$11 km) resolution grid cell for which daily rainfall time series
are extracted. The dense, near-continuous coverage across the subcontinent
illustrates the fine spatial resolution at which the subsequent classification is
performed.}
    \label{fig:gridpoints}
\end{figure}

In this study we used the GSMaP ISRO dataset obtained from the MOSDAC Open Data Portal~\cite{mosdac}. The dataset is available at a spatial resolution of $0.1^\circ$ ($\sim$11~km) and a temporal resolution of 1~hour. For the purpose of our study, we consider only the 29,026 grid points over the Indian landmass as shown in \cref{fig:gridpoints}. We accumulate the data into 24-hour windows to obtain daily rainfall data. We consider data from 2001 to 2022, excluding 2012, giving 21 years of observations. Due to data integrity issues encountered during processing, the GSMaP ISRO records for 2012 could not be reliably recovered and were excluded from the analysis. Further background on rainfall datasets and reanalysis products over India can be found in \cite{kumar2022,kumar2024}.

Studies have shown that GSMaP ISRO provides a reliable representation of daily
rainfall variability over the Indian landmass~\cite{kumar2025}. Satellite-based
rainfall products of this type have been validated against in-situ observations in
diverse tropical settings~\cite{macharia2022}, supporting their use as a primary
data source where dense gauge networks are unavailable. The
dataset is developed by applying corrections to the GSMaP precipitation estimates
using IMD's gridded rainfall dataset. GSMaP uses geostationary infrared (IR) and
microwave (MW) radiometer observations from multiple satellite sources to estimate
precipitation~\cite{kubota2020}.

Daily rainfall distributions are typically highly skewed, with a large number of
zero-rainfall events. To reduce the skewness and improve the suitability of the data for
statistical analysis, a logarithmic transformation is applied:
\begin{equation}
    \text{rainfall (in mm)} \longrightarrow \log(1 + \text{rainfall (in mm)} + \epsilon),
    \label{eq:logtransform}
\end{equation}

where $\epsilon = 10^{-2} \text{ mm}$ is a small bias term added to avoid numerical issues
associated with zero-rainfall events.

\begin{figure}[t]
    \centering
    \includegraphics[width=\linewidth]{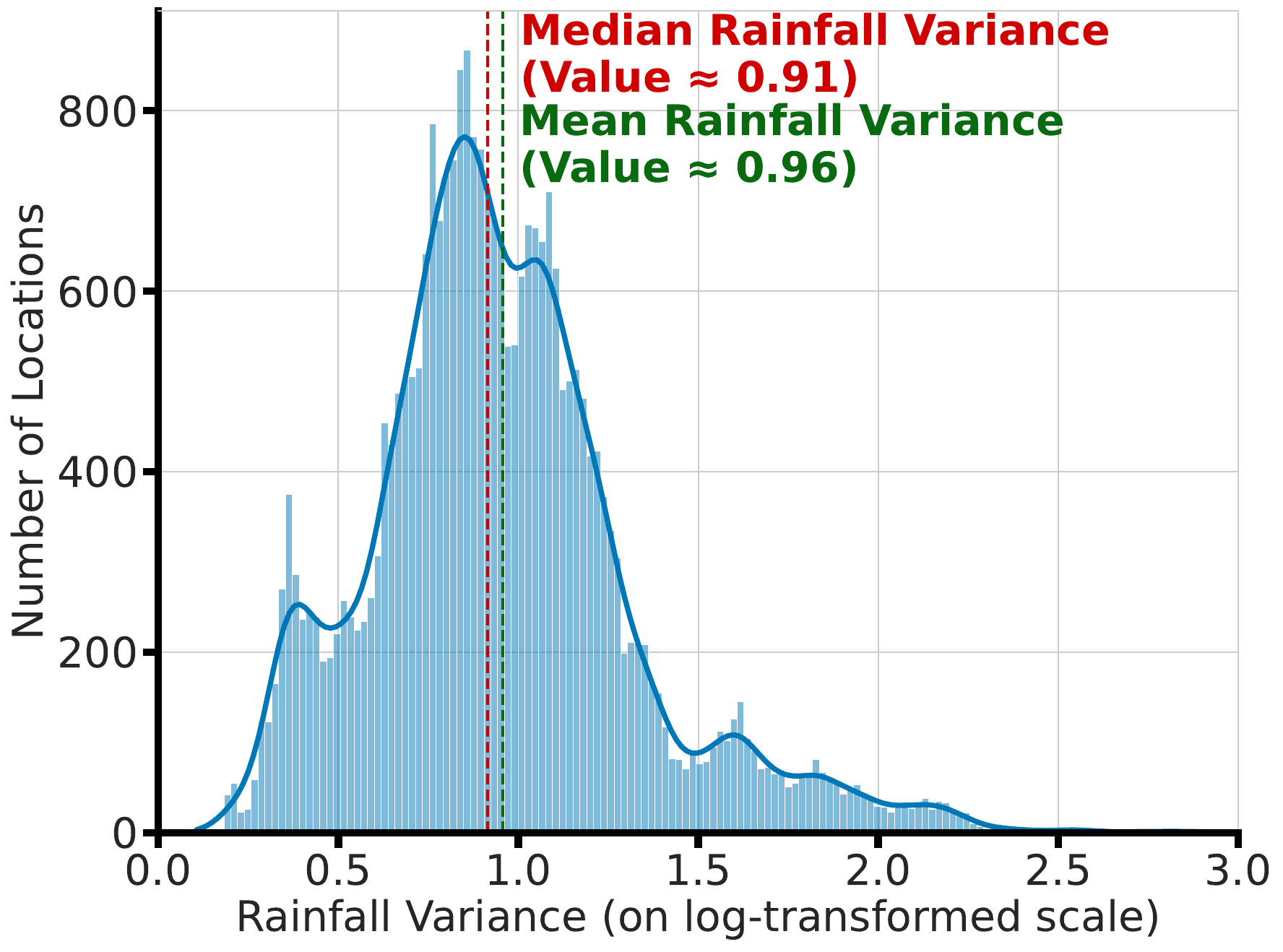}
    \caption{Distribution of the variance of Indian rainfall across all 29,026 grid
points. Histogram of the variance of log-transformed daily rainfall
($\epsilon = 10^{-2} \text{ mm}$), with the median rainfall variance (0.91) marked by the red dashed line and the mean rainfall variance (0.96) marked by the green dashed line.
The distribution is positively skewed, with most locations clustered near the median
and exhibiting moderate intra-annual variability, while a smaller subset of locations
extends into a long right tail of substantially higher variability.}
    \label{fig:variance_hist}
\end{figure}

\begin{figure}[t]
    \centering
    \includegraphics[width=\linewidth]{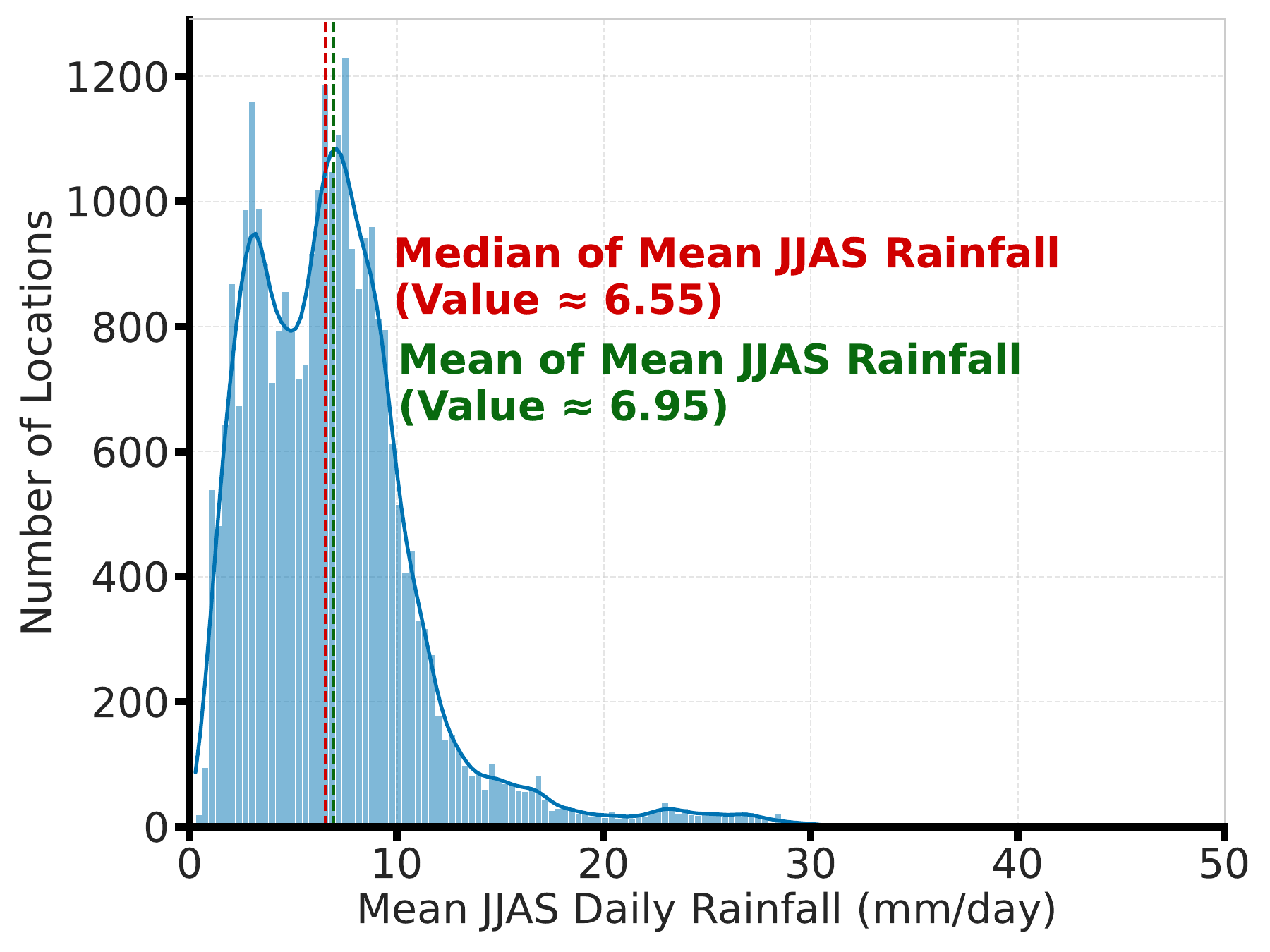}
    \caption{Distribution of mean JJAS rainfall of Indian rainfall across all 29,026 grid
points. Histogram of mean JJAS daily rainfall, with the median of the mean JJAS rainfall (6.55 mm/day) marked by the red dashed line and the mean of the mean JJAS rainfall (6.95 mm/day) marked by the green dashed line. The histogram shows a distinctly bimodal distribution with peaks near 2 and 8 mm/day, reflecting the coexistence of relatively dry and rainfall-rich
climatological regimes across the subcontinent. The pronounced right tail
corresponds to a small number of locations receiving exceptionally high monsoonal
rainfall.}
    \label{fig:jjas_hist}
\end{figure}

The variance histogram (\cref{fig:variance_hist}) shows a positively skewed distribution, with most grid points exhibiting rainfall variance values close to the median rainfall variance (0.91) and a relatively small fraction displaying substantially higher variance. This indicates that the intra-annual rainfall variability differs considerably among locations. The distribution of mean June-July-August-September (JJAS) rainfall (\cref{fig:jjas_hist}) is distinctly bimodal, with two prominent peaks near 2 and 8 mm/day. This suggests the presence of two dominant climatological rainfall regimes rather than a single continuous distribution, consistent with clustering-based studies that similarly identify two broad rainfall regimes across India \citep{rajak2026comparative, saha2018disparity}. In addition, the pronounced right tail reflects a relatively small number of locations experiencing exceptionally high monsoonal rainfall. Together, these distributions highlight the substantial diversity in rainfall characteristics across India and provide the climatological context for the graph-based analysis presented in the following sections.

\subsection{Categorizing Locations Based on Intra-Annual Rainfall Variability}
\label{subsec:categorize}

\begin{figure}[htb]
    \centering
    \includegraphics[width=\linewidth]{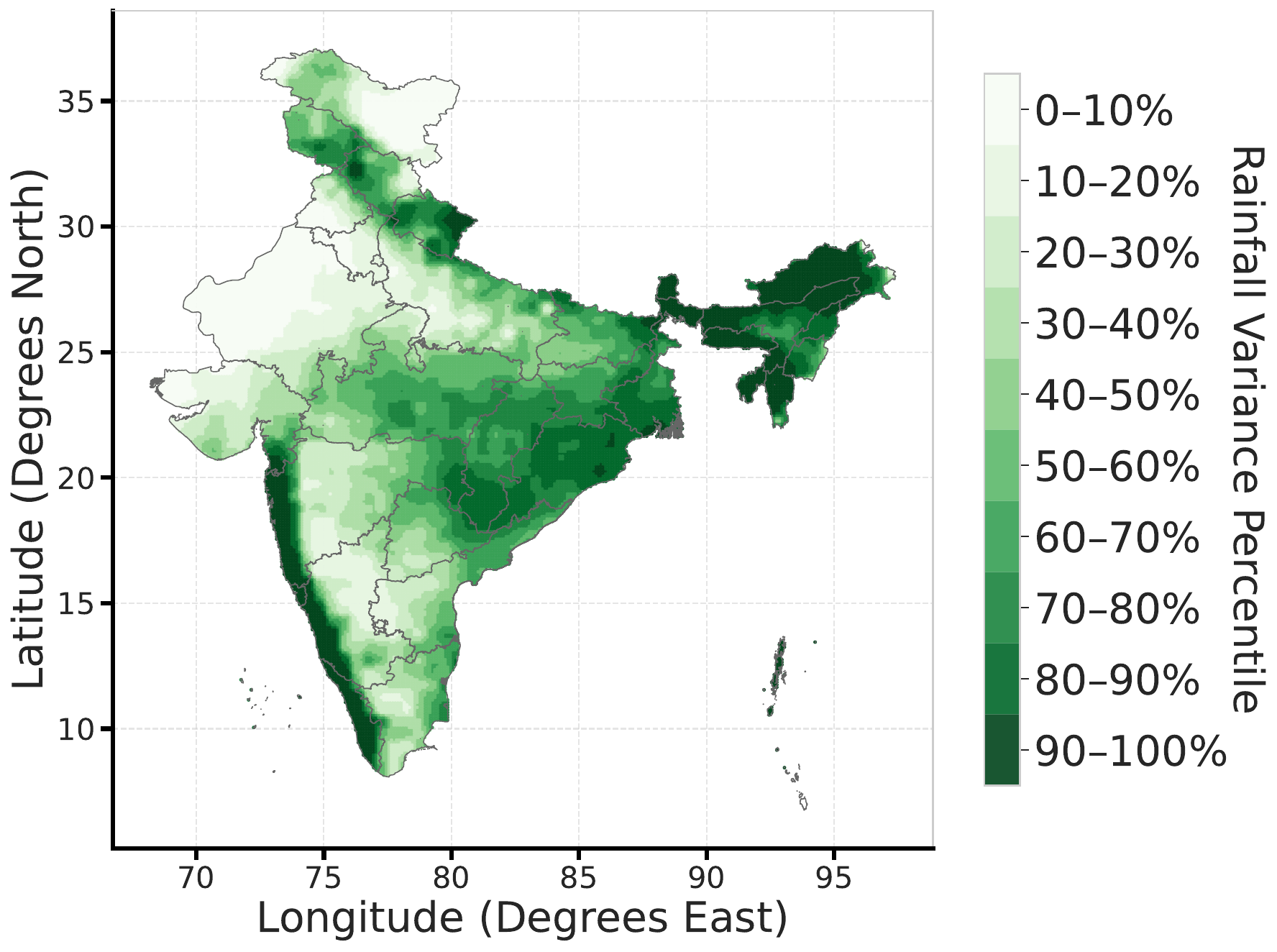}
    \caption{Spatial distribution of intra-annual rainfall variability across India,
with locations grouped into 10 discrete bins according to their percentile rank in
intra-annual variance. Darker shading indicates higher variability. The West Coast
and Northeast India stand out with the highest intra-annual variability, while
Western Rajasthan and Eastern Ladakh exhibit the lowest, illustrating the pronounced
spatial heterogeneity in day-to-day rainfall fluctuations across the subcontinent.}
    \label{fig:variance_map}
\end{figure}

The log-transformed rainfall distribution exhibits a pronounced right tail, indicating
substantial variability in rainfall intensity across locations. To investigate this
spatial heterogeneity, we categorize locations based on their intra-annual rainfall
variability, defined as the variance of daily rainfall within a given year
computed for each location and subsequently averaged across all years. Locations are
grouped into 10 discrete categories corresponding to their percentile rank in
intra-annual variability. The resulting spatial distribution is shown in
\cref{fig:variance_map}. As shown in \cref{fig:variance_map}, the West Coast and Northeast India exhibit the highest variability, while Western Rajasthan and Eastern Ladakh exhibit the lowest.

\subsection{Generating the Graphs}
\label{subsec:graphs}

The central methodological step of this study is the conversion of each location's
21-year rainfall record into a similarity network: individual years become nodes,
and the pairwise similarity between their daily rainfall trajectories becomes edge
weight. Rainfall consistency, a temporal property, is thereby re-expressed as network
topology, a structural property, which the GNN can then learn from directly.

For training, 200 locations are randomly selected from each of the 10 variability bins,
giving 2,000 locations in total. The choice of 200 locations per bin balances two competing objectives. First, equal sampling across bins ensures that all variability regimes, from the most arid to the most rainfall-intensive, contribute equally to training, preventing the model from being biased towards the mid-range bins that contain the majority of Indian grid points. Second, the total training set of 2,000 locations represents a small fraction ($\sim$6.9\%) of the full 29,026-point domain, reflecting the expectation that the graph representation encodes sufficient local structure for the model to generalize across the landmass without requiring exhaustive coverage of the spatial domain. This stratified sampling ensures representation across
the full spectrum of rainfall variability, and partially mitigates the class imbalance
that would arise from purely random sampling. This random selection was repeated across multiple independent trials with different
random seeds. The resulting model performance and spatial classification pattern
showed no meaningful sensitivity to the particular sample drawn, indicating that the
training procedure is robust to the specific random draw of locations rather than an
artifact of any one sample. Each location contains 21 years of daily
rainfall data transformed as per \cref{eq:logtransform}. February~29 is excluded from
all leap years, yielding 365 observations per year.

For each location, a graph is constructed in which individual years are represented as
nodes, each associated with a $365\times 1$ feature vector of daily rainfall values. A climatological node representing the time series of long-term mean rainfall (2001–2022) was added to each graph and connected to all annual rainfall nodes. This design ensures graph connectivity, facilitates information exchange between anomalous and typical years, and enables the GNN to jointly learn inter-annual similarity patterns and departures from climatological conditions.

Edges between year nodes are established based on cosine similarity between
corresponding rainfall vectors. For two vectors $\mathbf{r}_1$ and $\mathbf{r}_2$:
\begin{equation}
    \text{cosine similarity} = \frac{\mathbf{r}_1 \cdot \mathbf{r}_2}
                                    {|\mathbf{r}_1||\mathbf{r}_2|}.
    \label{eq:cosine}
\end{equation}
This metric captures the similarity in temporal rainfall patterns independent of
magnitude \citep{jones1987pictures}. Cosine similarity is conventionally bounded in
$[-1,1]$ for arbitrary real-valued vectors \citep{chanwimalueang2017cosine}. In the
present case, however, all feature vectors are constructed from log-transformed daily
rainfall values (\cref{eq:logtransform}), which are strictly non-negative by
construction. Consequently, both $\mathbf{r}_1$ and $\mathbf{r}_2$ lie entirely in the
non-negative orthant, so their dot product $\mathbf{r}_1 \cdot \mathbf{r}_2 \geq 0$,
and cosine similarity is restricted to the range $[0,1]$ rather than the full
$[-1,1]$ range possible for unconstrained vectors \citep{jones1987pictures,
korenius2007principal, sidorov2014soft}. For each location, pairwise similarities are
computed across all yearly nodes, yielding $\binom{21}{2} = 210$ values. Across all
training locations, this gives 420,000 similarity scores. With values near to 1
indicating stronger similarity, cosine similarity remains independent of the
magnitude of rainfall. High cosine similarity between two nodes indicates that the
seasonal evolution of rainfall at that grid point was similar in those two years,
implying comparable timing and progression of active and break periods, although not
necessarily similar rainfall magnitudes or seasonal totals.

Cosine similarity was preferred over alternatives such as dynamic time warping because
the log transformation guarantees strictly positive feature values, which bounds the
similarity in $[0, 1]$ and makes a global percentile threshold directly interpretable
as a minimum pattern similarity for edge formation.

A global threshold is defined as the 75\textsuperscript{th} percentile of all pairwise similarity values
computed on the training set. Edges are retained between node pairs whose similarity exceeds this threshold. The resulting graphs are treated as undirected. 

\subsection{Generating the Labels for Training}
\label{subsec:labels}

Each location is categorized as either consistent or erratic based on
the similarity of its annual rainfall patterns over time. A consistent location is
one whose daily rainfall evolution follows a broadly repeatable seasonal trajectory
from year to year, with active and break periods occurring at similar times and in
similar sequence across different years, even if the absolute rainfall magnitudes
vary. An erratic location, by contrast, is one whose year-to-year rainfall evolution
lacks such repeatability: the timing, sequencing, and progression of wet and dry
spells differ substantially from one year to the next, such that no single year's
daily rainfall pattern is strongly representative of another's. This distinction
characterizes the shape and timing of a location's seasonal rainfall
cycle across years, rather than its overall rainfall amount; a location can therefore
be classified as consistent regardless of whether it is climatologically wet or dry,
provided its year-to-year seasonal pattern is repeatable.

Operationally, this consistency is quantified using pairwise cosine similarity
(\cref{eq:cosine}) between each pair of yearly rainfall vectors at a location. The
mean of all pairwise cosine similarity values is computed for each location, giving a
single scalar measure of how self-similar that location's rainfall evolution has been
across the 21 years of the study period. A global threshold at the 75\textsuperscript{th} percentile of
these means is then applied. Locations above the threshold, i.e., those whose years are
on average more mutually similar than the majority of locations across India, are
labeled consistent, while the remainder are labeled erratic.

An important consequence of this labeling strategy is its direct relationship with
graph structure: consistent locations produce more densely connected graphs (more node
pairs exceed the edge threshold), while erratic locations yield sparser graphs. We note
that both the edge connectivity and the class labels are derived from the same pairwise
cosine similarity values. The physical validity of the resulting classification is
assessed in \cref{sec:validation}.

\subsection{Model and Training}
\label{subsec:model}

A GNN is employed to classify locations based on their rainfall variability patterns.
The architecture is implemented using PyTorch Geometric~\cite{fey2019fast} and consists of
two graph convolutional layers (GCNConv)~\cite{kipf2016semi} followed by a
global mean pooling readout and a fully connected classification layer. The first
convolutional layer maps the 365-dimensional input features to a 64-dimensional latent
space. The second layer refines these representations while maintaining the
64-dimensional space. The global mean pool aggregates node embeddings into a single
graph-level representation, which is then mapped to two output classes by a linear
layer. Global mean pooling is appropriate here because all nodes, year-nodes and the
climatological node alike, contribute meaningful signal to the graph-level
classification. Mean aggregation preserves this distributed information, whereas max
pooling would discard it~\cite{xu2018powerful}.

Dropout with rate 0.3 is applied after each convolutional layer and after the
readout step. Class weights are computed from the training label distribution and
incorporated into the cross-entropy loss to account for class imbalance. Training
uses the Adam optimizer with a learning rate of $5\times10^{-3}$ and a weight decay of
$10^{-3}$, with a learning rate scheduler (ReduceLROnPlateau, factor 0.5,
patience 5). The training set reserves 20\% for validation. The batch size is 128 with
a maximum of 120 epochs and early stopping (patience 10) based on validation
accuracy. The model converged after 38 epochs, achieving a training accuracy of
98.4\% and a validation accuracy of 96.8\%.

\begin{figure*}[t]
    \centering
    \includegraphics[width=\linewidth]{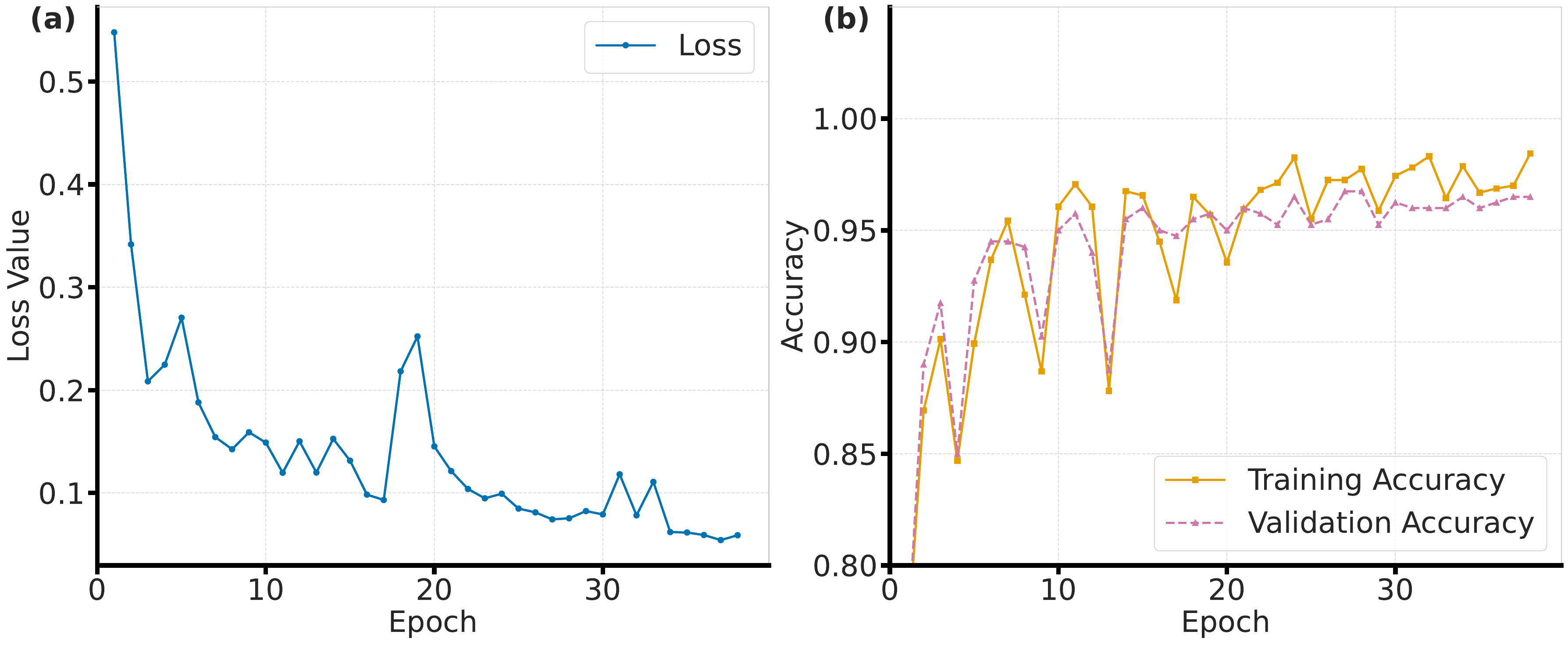}
    \caption{Training dynamics of the GNN over 38 epochs. (a) Loss optimization
history, showing a rapid decrease over the first 10 epochs before stabilizing near
0.05. The transient spikes around epochs 10--20 correspond to learning rate
reductions triggered by ReduceLROnPlateau. (b) Training and validation accuracy
across epochs, both exceeding 0.96 by the final epochs and tracking one another
closely throughout training, with no significant divergence between the two curves,
indicating an absence of overfitting. Early stopping was triggered at epoch 38.}
    \label{fig:loss}
\end{figure*}

\Cref{fig:loss} illustrates these training dynamics, showing a rapid initial decrease
in loss followed by stabilization, along with training and validation accuracy
curves that track one another closely, indicating stable convergence without
overfitting.

\begin{figure}[htb]
    \centering
    \includegraphics[width=\linewidth]{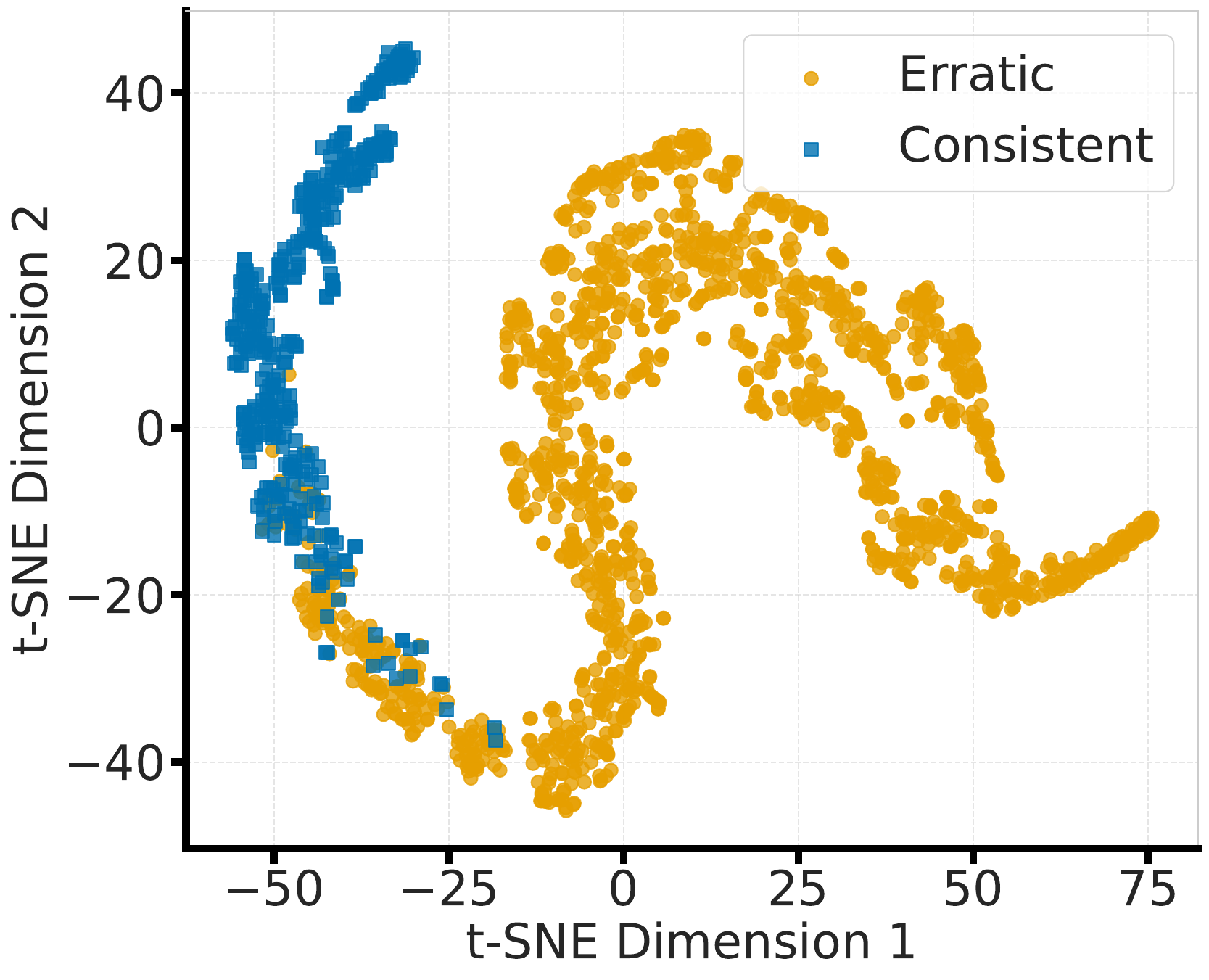}
    \caption{Two-dimensional t-SNE projection of the graph-level embeddings learned
by the trained GNN, with each point corresponding to one training location, colored
by predicted class (blue: consistent; yellow: erratic). The two classes form
distinct, well-separated clusters in the embedding space, indicating that the model
has learned a discriminative representation of rainfall consistency rather than
producing an arbitrary or overlapping partition of the graphs.}
    \label{fig:tsne}
\end{figure}

\Cref{fig:tsne} shows a t-SNE~\cite{van2008visualizing} projection of the graph-level
embeddings produced by the trained model. t-SNE (t-distributed Stochastic Neighbor
Embedding) is a dimensionality-reduction technique that maps high-dimensional data
into a low-dimensional (here, two-dimensional) space while preserving local
neighborhood structure, such that points that are similar in the original
high-dimensional space remain close together in the projection~\cite{van2008visualizing}.
It is used here purely for visualization, allowing the high-dimensional graph-level
embeddings learned by the GNN to be inspected qualitatively. The two classes are
well-separated in the embedding space (as shown in \cref{fig:tsne}), confirming that the model has learned a
discriminative representation of rainfall variability.

\section{Results}
\label{sec:results}

\subsection{Classification of the Regimes of Indian Rainfall}
\label{subsec:classification}

\begin{figure}[htb]
    \centering
    \includegraphics[width=\linewidth]{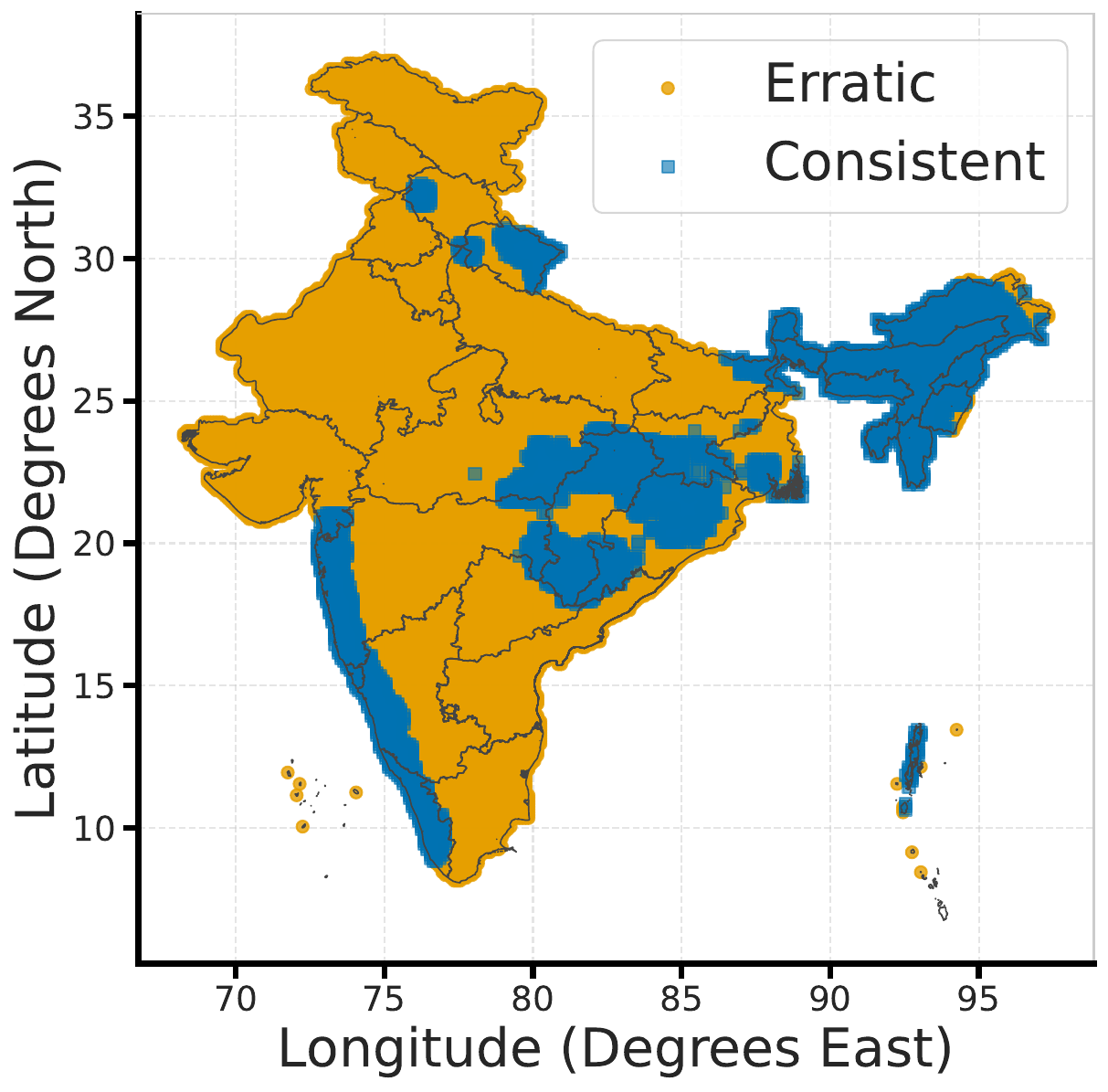}
    \caption{Spatially continuous classification of all 29,026 grid points across
India, produced by applying the trained GNN to the full landmass (blue: consistent;
yellow: erratic). Consistent locations are concentrated along the Western Ghats,
Northeast India, and parts of central India, coinciding with regions of early
monsoon onset and high seasonal rainfall, while erratic locations dominate the
northwest, the Deccan Plateau, and much of the Gangetic Plain.}
    \label{fig:predictions}
\end{figure}

The consistent regions identified by the model include the Western Ghats,
Northeast India, and parts of central India, as shown in \cref{fig:predictions}. The first two coincide with
regions of early monsoon onset and high seasonal rainfall, while parts of central
India are characterized by relatively persistent monsoonal rainfall. These
characteristics provide physical support for the identified consistent rainfall regime.

\begin{figure}[htb]
    \centering
    \includegraphics[width=\linewidth]{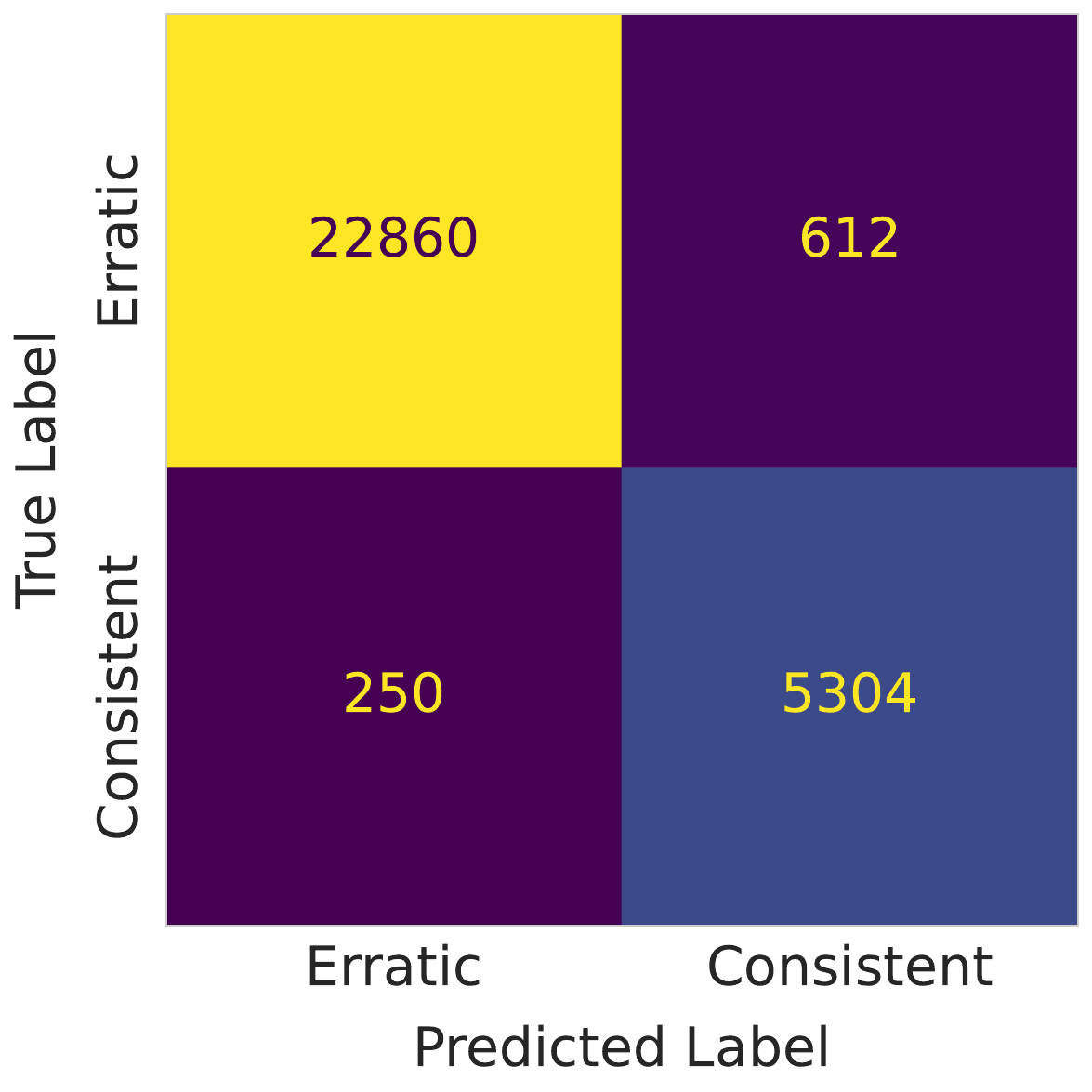}
    \caption{Confusion matrix summarizing GNN classification performance across all
29,026 grid points, comparing predicted class against the cosine-similarity-based
class label. The model correctly classifies 22,860 of 23,472 erratic locations
(97.4\% recall) and 5,304 of 5,554 consistent locations (95.5\% recall), with
misclassifications concentrated in the off-diagonal cells (612 erratic locations
predicted as consistent, and 250 consistent locations predicted as erratic),
indicating strong overall agreement between the model's predictions and the
underlying labels.}
    \label{fig:confusion}
\end{figure}

Classification performance is summarized using a confusion matrix
(\cref{fig:confusion}), a standard tool for evaluating classifier performance
that cross-tabulates the true class of each sample against the class predicted
by the model, allowing the number of correctly and incorrectly classified samples
in each category to be directly assessed \citep{tharwat2021classification}. From this
matrix, standard performance metrics such as precision, recall, and F1-score are
derived (\cref{tab:metrics}). The F1-score in particular is defined as the harmonic
mean of precision and recall, providing a single balanced measure of classifier
performance that is especially informative when class sizes are imbalanced, as is
the case here \citep{lipton2014optimal}. The results confirm strong classification
performance: out of the 23,472 erratic locations, 22,860 are correctly classified
(97.4\% recall), and of the 5,554 consistent locations, 5,304 are correctly classified
(95.5\% recall).

\begin{table}[H]
    \centering
    \caption{Performance metrics derived from the confusion matrix.}
    \label{tab:metrics}
    \begin{ruledtabular}
        \begin{tabular}{lcccc}
            \toprule
            \textbf{Class / Metric} & \textbf{Precision} & \textbf{Recall}
                                    & \textbf{F1-score}  & \textbf{Support} \\
            \hline
            Erratic          & 0.99 & 0.97 & 0.98 & 23{,}472 \\
            Consistent       & 0.90 & 0.95 & 0.92 &  5{,}554 \\
            \hline
            Accuracy         & ---  & ---  & 0.97 & 29{,}026 \\
            Macro Average    & 0.94 & 0.96 & 0.95 & 29{,}026 \\
            Weighted Average & 0.97 & 0.97 & 0.97 & 29{,}026 \\
            \bottomrule
        \end{tabular}
    \end{ruledtabular}
\end{table}

Both classes achieve an F1-score above 0.90, confirming strong model performance.
The consistent regions identified by the model, the West Coast, Northeast India,
and parts of central India, are among the first regions where the southwest monsoon
onset occurs and which receive the highest seasonal rainfall, making the
classification physically convincing.

\begin{figure*}[t]
    \centering
    \includegraphics[width=\linewidth]{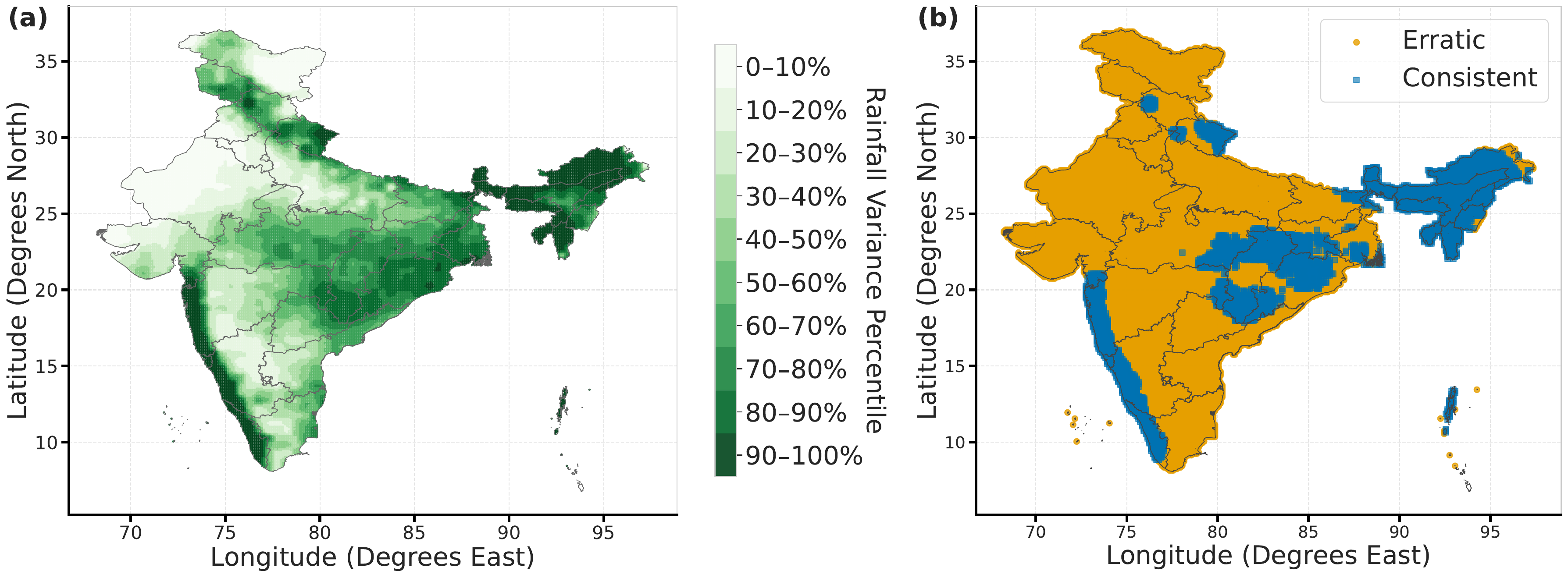}
    \caption{Side-by-side comparison of intra-annual rainfall variability and
inter-annual rainfall consistency across India. (a) Spatial distribution of
intra-annual rainfall variability, reproduced from \cref{fig:variance_map}. (b)
Spatially continuous classification of rainfall regimes identified by the GNN
(blue: consistent; yellow: erratic), reproduced from \cref{fig:predictions}. A close
spatial correspondence is evident between regions of high intra-annual variability
in (a) and locations classified as consistent in (b), particularly over the Western
Ghats and Northeast India, indicating a link between rainfall variability operating
at intra-annual and inter-annual timescales.}
    \label{fig:comparison}
\end{figure*}

The comparison between the spatial distribution of intra-annual rainfall variability
with the rainfall regimes identified by the GNN is shown in \cref{fig:comparison}. A strong spatial correspondence is evident between regions of high intra-annual rainfall
variability (dark green in \cref{fig:comparison}(a)) and locations classified as
consistent (blue in \cref{fig:comparison}(b)), particularly over the Western Ghats
and Northeast India. Conversely, regions exhibiting lower intra-annual rainfall
variability, including the Deccan Plateau and northwestern India, are predominantly
classified as erratic.

At first glance, this relationship may appear counterintuitive: one might expect
locations with larger day-to-day rainfall swings to also exhibit less predictable
behavior from year to year. However, intra-annual variance as defined here reflects
the amplitude of a location's seasonal rainfall cycle, that is, the contrast
between the wet monsoon season and the drier remainder of the year, rather than the
randomness or noisiness of its rainfall. A large seasonal amplitude typically
indicates that rainfall at that location is governed by a strong, physically robust
forcing mechanism. Over the Western Ghats, this forcing arises from orographic
uplift of the moisture-laden cross-equatorial low-level monsoon jet as it encounters
the mountain barrier \citep{halder2022dynamical, shige2017role, phadtare2022froude},
while over Northeast India, a comparable role is played by topographic funneling of
moisture around the Brahmaputra valley and Meghalaya Plateau, reinforced by
low-level-jet-driven moisture convergence from the Bay of Bengal
\citep{das2024dynamics, fujinami2017contrasting, prokop2015variation}. Such
large-scale, boundary-forced mechanisms tend to recur reliably from year to year,
in line with the broader finding that seasonal rainfall components tied to strong,
persistent large-scale forcing are inherently more reproducible than those governed
by smaller-scale, internally generated intraseasonal variability
\citep{shukla2006predictability, krishnamurthy2008seasonal, zhou2009well}. Locations
with a strong, dominant seasonal signal of this kind therefore tend to exhibit a
highly repeatable seasonal rainfall pattern across years, even though the
within-year swing is large. By contrast, locations with weak or muted seasonality
(low intra-annual variance) are more strongly influenced by smaller-scale, less
regular convective processes that are not tied to a dominant recurring forcing,
resulting in less repeatable year-to-year behavior. As shown earlier, regions of
high intra-annual rainfall variability are also climatologically rainfall-rich.
Taken together, these findings suggest that rainfall-rich regions, whose seasonal
cycle is dominated by a strong and physically robust forcing, generally exhibit
greater inter-annual consistency in their annual rainfall patterns than
rainfall-deficient regions, where weaker and more localized forcing produces less
repeatable behavior.

\section{Validation}
\label{sec:validation}

\subsection{Statistical Validation Against Physical Metrics}
\label{subsec:stats}

To verify that the GNN classification captures physically meaningful differences, we
compute three statistical metrics for each of the 29,026 locations and compare their
distributions between the consistent and erratic classes: (i)~intra-annual rainfall
variance, defined as the variance of daily rainfall within a given year averaged across
all years, (ii)~intra-annual coefficient of variation (CV), defined as the ratio of the
standard deviation of intra-annual rainfall to its mean across years, and
(iii)~mean number of rainy days per year, where a rainy day is defined as a day with
rainfall exceeding 1~mm.

\begin{figure*}[htb]
    \centering
    \includegraphics[width=\linewidth]{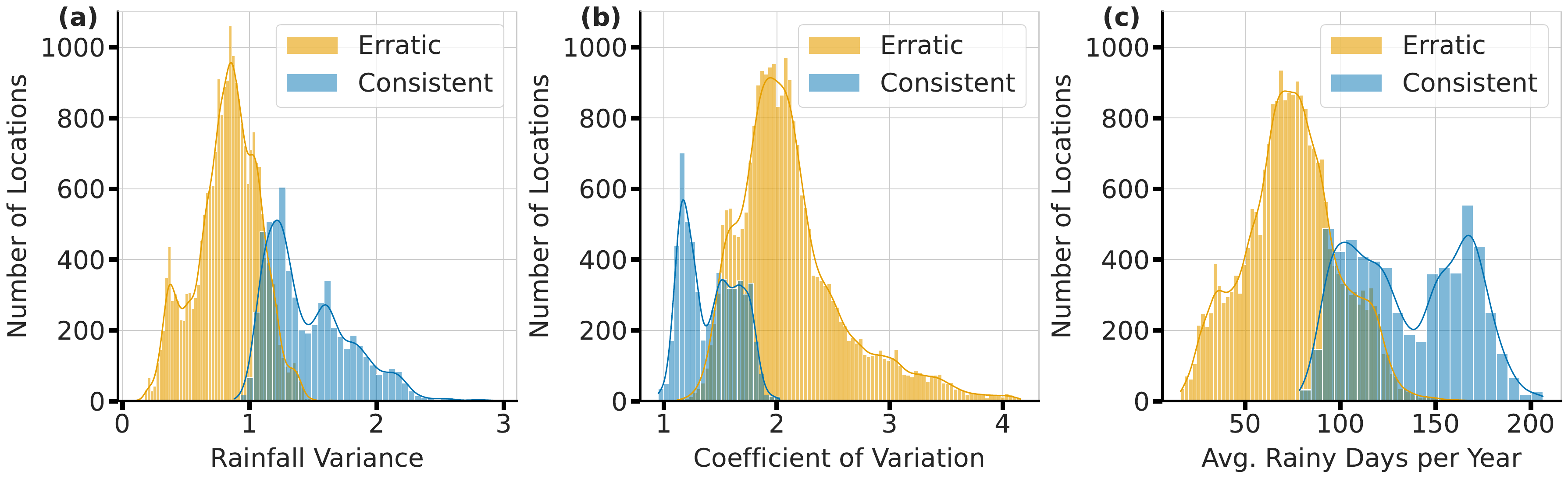}
    \caption{Distributions of three independent statistical metrics, stratified by
GNN-predicted class (blue: consistent; yellow: erratic), used to validate the
classification against physical climatological characteristics not used during
training. (a) Intra-annual rainfall variance, with consistent locations exhibiting a
broader distribution (relative to the number of locations) skewed toward higher values than erratic locations.
(b) Inter-annual coefficient of variation (CV), with erratic locations concentrated
at higher CV values and a longer tail, and consistent locations clustered at lower
values. (c) Mean number of rainy days per year, showing the most pronounced
separation between classes, with consistent locations experiencing substantially
more rainy days than erratic locations. A Mann--Whitney U test confirms that the two
classes differ significantly for all three metrics ($p \approx 0$ in each case).}
    \label{fig:metrics}
\end{figure*}

The separation between the two classes is clearly visible across all three metrics
in \cref{fig:metrics}. In panel (a), consistent locations exhibit a broader
distribution (relative to the number of locations) of intra-annual rainfall variance
extending to higher values than erratic locations, indicating stronger intra-annual
rainfall variability. This is corroborated in panel (c), where consistent locations
also show a substantially greater number of rainy days than erratic locations,
together pointing to more persistent rainfall occurrence. In panel (b), by contrast,
erratic locations are concentrated at higher CV values, with a longer tail, implying
greater rainfall irregularity and a more sporadic, episodic occurrence of rainfall
relative to the local mean rainfall. Although consistent regions experience larger
absolute rainfall fluctuations (panel a), their lower CV (panel b) indicates that
these fluctuations remain moderate relative to the higher climatological mean
rainfall. The distinction is most pronounced in the rainy-day distributions (panel
c), where consistent regions typically experience approximately 150--175 rainy days
per year compared with about 50--80 days in erratic regions. This suggests that
consistent regions receive rainfall over a substantially larger fraction of the
annual cycle, whereas rainfall in erratic regions is concentrated into fewer, more
episodic events. Collectively, these results highlight a fundamental distinction
between rainfall-rich and rainfall-deficient regions of the Indian subcontinent: the
former exhibit larger absolute rainfall fluctuations but lower relative variability,
whereas the latter display smaller absolute fluctuations yet greater relative
variability.

To assess whether these differences are statistically significant rather than due to
chance, a Mann--Whitney U test is applied to each metric. The Mann--Whitney U test is a
non-parametric, distribution-free statistical test that compares the distributions of
two independent samples without assuming normality, making it well suited to the
non-Gaussian, positively skewed distributions of rainfall-derived metrics observed
here \citep{nachar2008mann}. Full details of the test procedure, including the null
hypothesis, test statistic, and significance criterion, are provided in
\cref{app:mannwhitney}. The test confirms that the distributions differ
significantly between the two classes for all three metrics ($p \approx 0$ in each
case), ruling out chance as an explanation. These results establish that the GNN
classification aligns with independent climatological characteristics of the
locations, providing strong evidence that the model captures genuine physical
structure rather than mathematical artifacts of the graph construction.

\subsection{Graph Structural Analysis}
\label{subsec:graph_metrics}

To complement the graph-based classification, we compute three graph-level structural
metrics for each location: average node degree, clustering coefficient, and graph density.
These metrics quantify the connectivity structure of the rainfall similarity graphs and provide
additional insights into the inter-annual consistency of rainfall patterns at each location.

\begin{figure*}[htb]
    \centering
    \includegraphics[width=\linewidth]{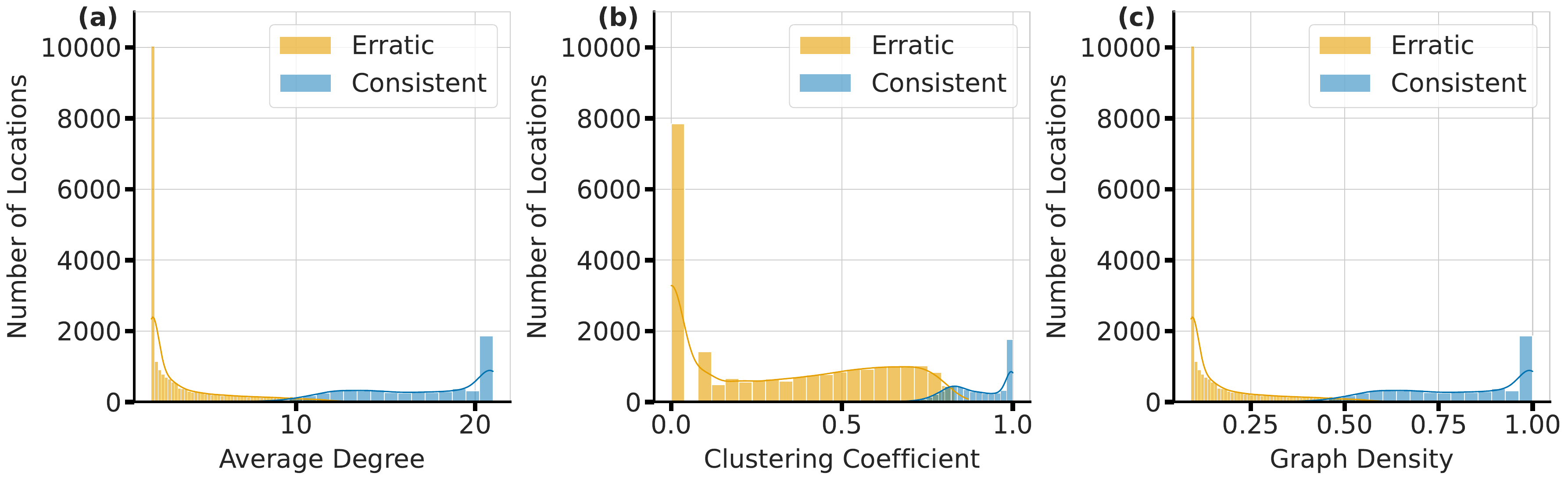}
    \caption{Distributions of three graph-level structural metrics, stratified by
GNN-predicted class (blue: consistent; yellow: erratic), quantifying the
connectivity of the rainfall similarity graphs independently of the class labels
themselves. (a) Average node degree. (b) Clustering coefficient. (c) Graph density.
In all three panels, consistent locations concentrate near the maximum possible
value, indicating highly interconnected graphs in which most years exhibit similar
rainfall evolution, whereas erratic locations predominantly exhibit low degree, low
density, and a broader spread of clustering coefficients, reflecting weaker
inter-annual connectivity. A Mann--Whitney U test confirms that the two classes
differ significantly for all three metrics ($p \approx 0$ in each case).}
    \label{fig:graph_metrics}
\end{figure*}

\Cref{fig:graph_metrics} shows clear separation between the two classes across all three metrics. Consistent locations are concentrated near the maximum possible average node
degree, clustering coefficient, and graph density, whereas erratic locations
predominantly exhibit low degree, low density, and a broader range of clustering
coefficients. This indicates that rainfall similarity graphs associated with consistent
locations are highly interconnected, with most years exhibiting similar rainfall
evolution patterns, while erratic locations produce much sparser graphs with weaker
inter-annual connectivity. A Mann--Whitney U test confirms statistically significant differences between the two classes for all three metrics ($p \approx 0$).

We note that this result is partially a consequence of the labeling strategy: since both labels and edges are derived from the same cosine similarity values, consistent locations are expected to exhibit higher mean similarity and consequently denser graph structures. Nevertheless, the clear separation observed in \cref{fig:graph_metrics} demonstrates that the graph representation effectively captures the underlying rainfall variability signal in a structured and discriminable manner. In particular, the distinction between the two rainfall regimes is reflected in globally dense graph connectivity rather than isolated local connections, providing a principled motivation for the use of GNNs in this classification task.

\subsection{Threshold Sensitivity Analysis}
\label{subsec:threshold}

Both the edge threshold and the labeling threshold are defined as percentiles of the
global cosine similarity distribution. To assess robustness to this choice, we vary
the threshold percentile from 70\% to 90\% and compare the resulting direct cosine
similarity classifications against the GNN predictions at 75\%.

\begin{table*}[htb]
    \centering
    \caption{Percentage of locations changing class relative to the GNN predictions
             at the 75\textsuperscript{th} percentile threshold, for varying threshold values.}
    \label{tab:threshold}
    \begin{ruledtabular}
        \begin{tabular}{cccc}
            \textbf{Threshold Percentile} & \textbf{Threshold Value} &
            \textbf{Locations Changed} & \textbf{\% Changed} \\
            \hline
            70\textsuperscript{th} & 0.5275 & 2{,}020 & 7.0\% \\
            75\textsuperscript{th} & 0.5502 & 887 & 3.1\% \\
            80\textsuperscript{th} & 0.5743 & 1{,}576 & 5.4\% \\
            90\textsuperscript{th} & 0.6348 & 3{,}662 & 12.6\% \\
        \end{tabular}
    \end{ruledtabular}
\end{table*}

\begin{figure*}[ht]
    \centering
    \includegraphics[width=\linewidth]{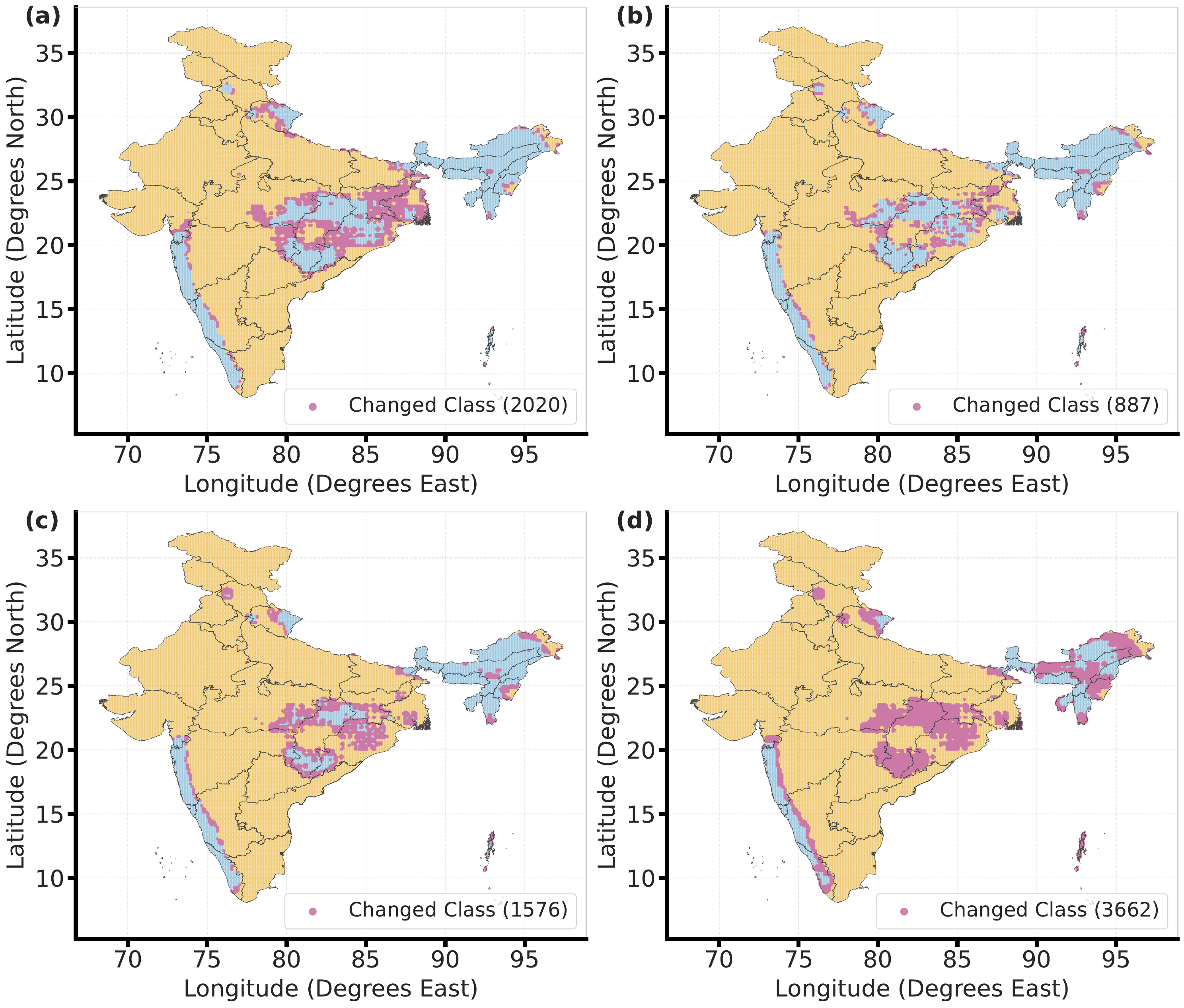}
    \caption{Spatial sensitivity of the GNN classification to the cosine similarity
edge and labeling threshold, evaluated by comparing direct cosine-similarity
classifications at each percentile against the GNN predictions obtained using the
baseline 75\textsuperscript{th} percentile threshold (blue: consistent; yellow: erratic; pink: changed
class relative to the baseline). (a) 70\textsuperscript{th} percentile (threshold = 0.5275; 2,020
locations changed). (b) 75\textsuperscript{th} percentile (threshold = 0.5502; 887 locations changed).
(c) 80\textsuperscript{th} percentile (threshold = 0.5743; 1,576 locations changed). (d) 90\textsuperscript{th}
percentile (threshold = 0.6348; 3,662 locations changed). Across all four panels,
changed locations are concentrated along the boundaries between consistent and
erratic regions rather than within their cores, indicating that the classification
is robust to reasonable variation in the threshold and that the affected locations
represent genuine transition zones between climatological regimes.}
    \label{fig:threshold}
\end{figure*}

As shown in \cref{tab:threshold}, varying the threshold by $\pm5$ percentile points
around the 75\textsuperscript{th} percentile mark changes the classification of only approximately 8\% of locations. Even
at the 90\textsuperscript{th} percentile, only 12.6\% of locations change class. \Cref{fig:threshold}
shows the spatial distribution of locations that change class at each threshold.
Critically, the changed locations (pink) are almost entirely concentrated along the
boundaries between consistent and erratic regions, while the cores of both classes
remain stable. Even at the most restrictive threshold (90\textsuperscript{th} percentile), where the number of changed locations is greatest, the large-scale spatial organization of the two rainfall regimes remains largely unchanged. This demonstrates the robustness of the classification to threshold selection and suggests that the affected locations represent transition zones between the two climatological regimes rather than fundamental changes in the large-scale spatial pattern.

It is worth noting that the GNN predictions used as the baseline in \cref{tab:threshold}
are not equivalent to a direct cosine similarity classification at the 75\textsuperscript{th} percentile.
The GNN is trained on 2,000 locations and learns to generalize the graph-structural
signal beyond the raw similarity values. It therefore produces a smooth decision
boundary that differs from a hard threshold applied directly to mean cosine similarity.
Critically, this means that even a direct cosine-similarity classification computed at
the 75\textsuperscript{th} percentile used to define the baseline does not perfectly match
the GNN's predictions. As shown in \cref{fig:threshold}(b), 887 locations are
classified differently by the two methods at this shared percentile. This nonzero
disagreement at the 75\textsuperscript{th} percentile itself is expected and is not an error, since the
two classifications are produced by fundamentally different decision rules, a smooth,
learned graph-level boundary in the case of the GNN, versus a hard threshold applied
directly to mean pairwise cosine similarity, rather than by the same rule evaluated
twice. The sensitivity analysis in \cref{tab:threshold} therefore captures two effects
simultaneously as the threshold percentile is varied away from 75\%: the genuine
movement of locations across the class boundary as the threshold changes, and this
same baseline residual disagreement between the GNN's learned boundary and the hard
cosine threshold, which persists at every percentile tested, including 75\% itself.
The spatial concentration of changed locations at region boundaries in
\cref{fig:threshold} suggests that the dominant effect as the threshold moves away
from 75\% is the former: these are genuinely ambiguous locations whose classification
is sensitive to the precise definition of consistency, rather than artifacts of the
GNN's generalization.

This discrepancy also clarifies the role of the GNN relative to a direct percentile
threshold on mean cosine similarity. Since computing the classification threshold
directly is computationally inexpensive even across the full 29,026-location domain,
the GNN is not motivated by computational efficiency. Rather, its purpose is
architectural: it operates on the full topology of each location's similarity
network, i.e., the year-wise rainfall trajectories and their pairwise connectivity,
rather than the single scalar mean similarity that defines the threshold rule. The
887-location disagreement reported above is direct evidence that the GNN has
learned something beyond this scalar summary. More fundamentally, the GNN's ability
to generalize, trained on only 2,000 locations yet matching independent
climatological metrics across all 29,026 unseen locations (\cref{sec:validation}),
is itself a validation that graph topology encodes a learnable, physically
meaningful signal. A static threshold rule offers no equivalent generalization to
validate. This also makes the GNN the more extensible foundation: it can
incorporate additional graph attributes, such as teleconnection indices, or
generalize to other hydroclimatic variables and datasets, in ways a fixed
percentile threshold cannot support.

\subsection{Temporal Stability Analysis}
\label{subsec:temporal}

To assess whether the classification is stable over time, we split the 21-year time
series into two periods, Period~1 (2001--2011) and Period~2 (2013--2022), and
compute cosine similarity classifications for each period using a single global
threshold (0.5569), obtained from the full 21-year dataset, applied consistently to
both periods. Using a common threshold, rather than independently re-estimating it
for each period, ensures that any observed disagreement between periods reflects
genuine changes in rainfall pattern classification rather than differences in the
threshold itself.

\begin{table*}[htb]
    \centering
    \caption{Inter-period classification agreement using a common cosine similarity
             threshold.}
    \label{tab:temporal}
    \begin{ruledtabular}
        \begin{tabular}{lc}
            \textbf{Threshold value (both periods)} & 0.5569 \\
            \hline
            \textbf{Inter-period agreement} & 27{,}161 / 29{,}026 locations (\textbf{93.6\%}) \\
        \end{tabular}
    \end{ruledtabular}
\end{table*}

\begin{figure*}[htb]
    \centering
    \includegraphics[width=\linewidth]{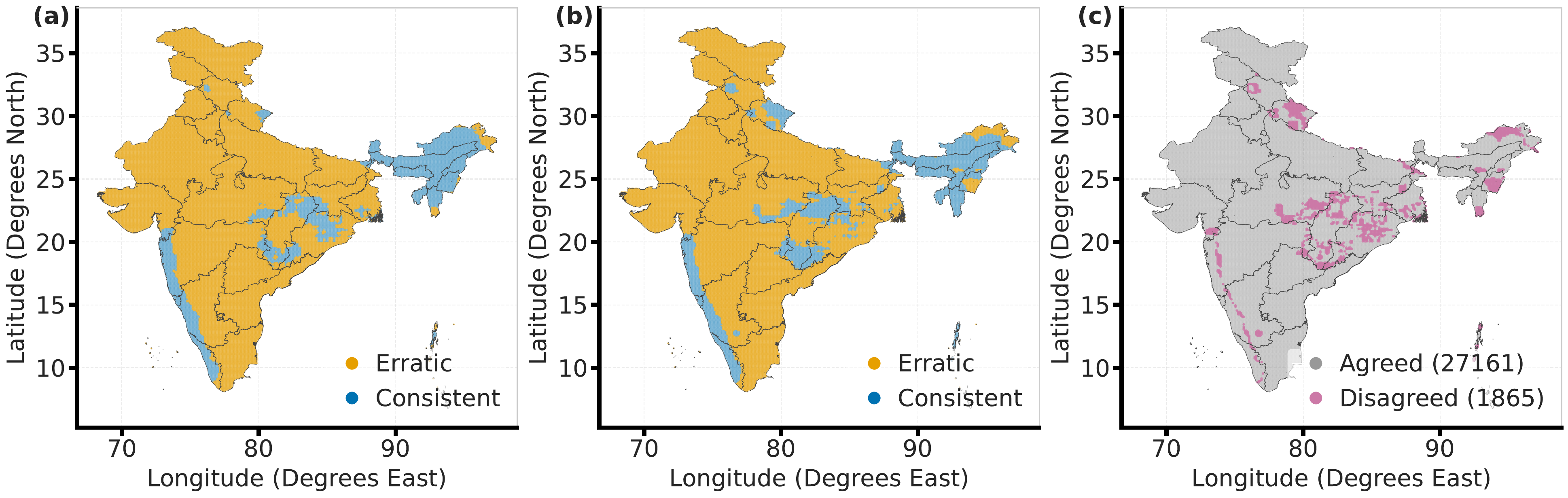}
    \caption{Temporal stability of the GNN classification across two independent study
periods. (a) Classification for Period 1 (2001–2011). (b) Classification for Period 2
(2013–2022). Consistent and erratic locations are shown in blue and yellow,
respectively. (c) Inter-period agreement between the two classifications. Gray
indicates locations assigned to the same class in both periods, whereas pink
denotes locations with differing classifications (27,161 agreement; 1,865
disagreement, corresponding to 93.6\% agreement).}
    \label{fig:temporal}
\end{figure*}

Classification agreement between the two periods is 93.6\% (\cref{tab:temporal}).
\Cref{fig:temporal} shows that the spatial classification patterns for the two
periods are nearly identical. The relatively few disagreements are concentrated
primarily along the transition boundaries separating the consistent and erratic
rainfall regimes, particularly near the eastern flank of the Western Ghats, the
western fringe of Northeast India, and parts of central India. In contrast, the
cores of both the consistent and erratic regions remain stable throughout the study
period.

The residual $\sim6.4\%$ disagreement may partly reflect genuine decadal climate variability
rather than classification uncertainty. For example, the 2001--2011 period includes
several strong La Ni\~na years, including the 2007/2008 and 2010/2011 events
\citep{li2022historical}, which are known to enhance the strength and expand the
spatial extent of the ISM \citep{xavier2007objective}. Consequently, at least part of
the observed disagreement may represent meaningful climatic variability rather than
noise. The persistence of core rainfall regimes across two independent time periods
further suggests that the identified spatial patterns represent stable
climatological structures rather than features specific to a particular analysis
period.

\section{Discussion}
\label{sec:discussion}

One of the principal findings of this study is the close spatial correspondence
between regions exhibiting high intra-annual rainfall variability and those classified
by the proposed GNN as inter-annually consistent. At first glance, this relationship
may appear counterintuitive: one might expect locations with larger day-to-day
rainfall swings to also exhibit less predictable behavior from year to year. However,
intra-annual variance as defined here reflects the amplitude of a location's
seasonal rainfall cycle rather than its randomness. A large seasonal amplitude
typically indicates rainfall governed by a strong, physically robust forcing
mechanism, such as orographic uplift of the cross-equatorial monsoon jet over the
Western Ghats \citep{halder2022dynamical, shige2017role, phadtare2022froude}, or
topographic moisture funneling and low-level-jet convergence over Northeast India
\citep{das2024dynamics, fujinami2017contrasting, prokop2015variation}. Such
large-scale, boundary-forced mechanisms recur reliably from year to year, consistent
with the broader finding that rainfall tied to strong, persistent large-scale forcing
is more reproducible than that governed by smaller-scale, internally generated
intraseasonal variability \citep{shukla2006predictability, krishnamurthy2008seasonal,
zhou2009well}. Locations with such a dominant seasonal signal therefore tend to
exhibit a highly repeatable seasonal rainfall pattern across years, even though the
within-year swing is large.

The results confirm this expectation: climatologically rainfall-rich regions,
particularly the Western Ghats, Northeast India, and parts of central India, exhibit
strong intra-annual rainfall variability while maintaining comparatively consistent
annual rainfall patterns across years. Conversely, rainfall-deficient regions are
characterized by lower intra-annual rainfall variability but greater relative
variability, that is, a higher coefficient of variation (CV), the ratio of the
standard deviation of intra-annual rainfall to its mean across years
(\cref{subsec:stats}). A high CV indicates that whatever rainfall these regions do
receive is distributed unevenly relative to their own (typically low) climatological
mean, producing a more sporadic, episodic occurrence of rainfall and, in turn, more
erratic inter-annual behavior. These findings suggest that rainfall variability
operating at intra-annual and inter-annual time scales is not independent, and that
rainfall-rich monsoon regions, whose seasonal cycle is dominated by a strong and
physically robust forcing, generally exhibit greater long-term consistency than
rainfall-deficient regions, where weaker and more localized forcing produces smaller
absolute fluctuations but greater relative variability and less repeatable behavior.

Beyond the specific classification presented here, the validation framework itself
constitutes a methodological contribution. Rather than relying on a single accuracy
metric, we assess physical validity through four largely independent lenses:
agreement with climatological metrics not seen during training, internal
graph-structural consistency, robustness to threshold choice, and stability across
disjoint time periods. This layered approach provides a general template for
validating graph-based classifications of environmental time series, applicable
beyond rainfall to other hydroclimatic or spatiotemporal variables where a
data-driven grouping must be shown to reflect genuine physical structure rather than
an artifact of graph construction or model capacity.

An important aspect of the proposed framework is its graph-based representation of
year-wise daily rainfall time series. Each node in the graph represents the daily
rainfall time series for one year at a given location, while edges connect pairs of
years whose rainfall evolution is sufficiently similar according to the cosine similarity
metric. Rather than treating individual years independently, the GNN exploits these
inter-year relationships to learn graph-level representations associated with rainfall
consistency. The clear separation between consistent and erratic locations in terms
of graph connectivity, together with the strong agreement with independent
climatological metrics, demonstrates that the graph representation captures
physically meaningful characteristics of rainfall variability rather than merely abstract
mathematical relationships. As shown in \cref{subsec:threshold}, this is not merely a matter of computational
convenience: the GNN's classification, learned from graph topology rather than a
single similarity scalar, measurably diverges from a direct threshold rule even at
matched percentiles, underscoring that the graph representation captures
information beyond what a hand-designed threshold can access.

The robustness analyses further strengthen confidence in the proposed framework.
The classification remains largely unchanged under moderate variations in the
cosine similarity threshold, with changes confined primarily to transition zones
between the two rainfall regimes while the cores of both classes remain stable.
Similarly, independent analyses of two decade-long sub-periods produce highly
consistent spatial classifications, suggesting that the identified rainfall regimes
represent persistent climatological structures rather than artifacts of a particular
analysis period. Collectively, these results indicate that the proposed framework is
robust to reasonable variations in graph construction and temporal sampling.

It is worth emphasizing that the large-scale spatial organization of the two
rainfall regimes is emergent rather than imposed. The GNN is trained on only 2,000
locations, selected without any explicit geographic information, and the model
architecture contains no spatial coordinates, adjacency, or prior notion of
contiguous regions. The coherent, spatially contiguous structure evident in
\cref{fig:predictions}, with the Western Ghats, Northeast India, and central India
forming continuous consistent regions rather than a scattered, spatially random
pattern, therefore arises purely from the local temporal similarity structure
learned independently at each grid point. This supports the interpretation that the
identified regimes reflect genuine, spatially coherent climatological structure
rather than an artifact of the classification procedure.

A methodological limitation of the present study is that both the graph edges and the
class labels are derived from the same pairwise cosine similarity values.
Consequently, the graph structure and the target labels are not entirely independent,
which may partially contribute to the observed classification performance.
Nevertheless, the close agreement between the GNN predictions and multiple
independent climatological metrics, together with the threshold sensitivity and
temporal stability analyses, suggests that the model captures physically meaningful
rainfall regimes rather than merely exploiting this structural dependence. This argument is strengthened by the fact that the classification's agreement with
external climatological metrics (\cref{subsec:stats}) is not merely statistically
significant but substantively large, evident in the direct visual separation shown
in \cref{fig:metrics}, and that this agreement persists under both threshold
perturbation (\cref{subsec:threshold}) and complete resampling into disjoint time
periods (\cref{subsec:temporal}). A purely circular signal, arising only from the
shared dependence of edges and labels on cosine similarity, would not be expected to
generalize this robustly across independent metrics, threshold choices, and time
periods that never enter the labelling procedure. Future work
should investigate alternative graph construction and labeling strategies that further
decouple edge generation from class definition. Incorporating additional climatic
information, such as large-scale teleconnection indices (e.g., ENSO and the Indian
Ocean Dipole), alternative similarity measures, or other hydroclimatic variables, may
further improve the physical interpretability and general applicability of graph-based
rainfall classification.

\section{Conclusion}
\label{sec:conclusion}

We reframed the problem of characterizing Indian monsoon rainfall consistency as one
of converting temporal variability into network topology. At each location, 21 years
of daily rainfall are organized into year-wise trajectories, whose pairwise
similarity defines a network whose connectivity and density directly encode how
repeatably that location's rainfall pattern recurs from year to year. A GNN trained
on this topology, rather than on raw rainfall values, classifies locations across
India as either inter-annually consistent or erratic. The resulting classification
identifies the Western Ghats, Northeast India, and parts of central India as
consistent rainfall regimes, in agreement with the known climatology of the Indian
summer monsoon. Notably, this large-scale spatial organization is emergent rather
than imposed: the model is trained on a small, geographically uninformed sample of
locations with no explicit spatial coordinates or adjacency structure, yet the
resulting classification coalesces into spatially coherent, contiguous regions
across the full Indian landmass. Independent validation using climatological
metrics, graph-structural characteristics, threshold sensitivity analysis, and
temporal stability across the 2001--2011 and 2013--2022 periods demonstrates that
the identified regimes are physically meaningful and robust, and together constitute
a general template for validating graph-based classifications of environmental time
series beyond the specific case of rainfall.

The proposed framework also reveals an unexpected coupling between
rainfall variability operating at different temporal scales. Locations with large
intra-annual rainfall amplitude, driven by strong, physically robust seasonal
forcing such as orographic uplift over the Western Ghats or moisture convergence
over Northeast India, also exhibit greater inter-annual consistency in their annual
rainfall patterns, whereas locations with weaker, more diffuse forcing exhibit both
smaller intra-annual amplitude and more erratic year-to-year behavior. This
indicates that intra-annual and inter-annual rainfall variability are not
independent properties but are linked through the strength and persistence of the
underlying physical forcing, a relationship that emerges directly from the network
topology without being explicitly modeled. To our knowledge, this study represents
one of the first applications of a similarity-network and graph classification
approach to characterize inter-annual rainfall consistency at grid-point scale
across the Indian landmass, linking graph-theoretic representations of temporal
rainfall similarity with physically interpretable climatological regimes.

The identified rainfall regimes have potential applications in regional water
resource management, agricultural planning, and climate risk assessment by
distinguishing areas characterized by relatively stable annual rainfall patterns
from those exhibiting greater inter-annual variability. More broadly, this study
demonstrates that converting year-wise time series into similarity networks, and
learning directly from their topology, is a viable and physically interpretable
strategy for characterizing long-term hydroclimatic variability. This approach
provides a foundation for extending graph-based topological analysis to other
hydroclimatic variables, satellite precipitation products, climatic regions, and
related climate classification problems. Future work could investigate alternative
graph construction and labeling strategies, incorporate large-scale climate drivers
such as ENSO and the Indian Ocean Dipole as additional graph attributes, and extend
the framework to sub-seasonal rainfall characteristics and other graph-based climate
applications.

\begin{acknowledgments}
\label{sec:acknowledgments}

This work was funded by IoE-IITM Research Initiatives (SP/22-23/1222/CPETWOCTSHOC) grant provided to R. I. Sujith.

\end{acknowledgments}

\section*{Data Availability Statement}
\label{sec:data_availability_statement}

The GSMaP ISRO precipitation data used in this study is publicly available through the Meteorological and Oceanographic Satellite Data Archival Centre (MOSDAC) of the Indian Space Research Organization at \url{https://www.mosdac.gov.in/opendata/GSMaP_SAC_RAIN}. The code used for data processing, analysis, and implementation of the graph neural network model will be made publicly available through a GitHub repository upon publication of this article.

\appendix
\section{Mann--Whitney U Test Procedure}
\label{app:mannwhitney}

For each of the statistical and graph-structural metrics reported in
\cref{subsec:stats,subsec:graph_metrics}, we test whether the distribution of that
metric differs significantly between the consistent and erratic classes using a
Mann--Whitney U test~\citep{nachar2008mann}.

Given two independent samples of size $n_1$ (consistent locations) and $n_2$
(erratic locations), the null and alternative hypotheses are: \\
\newline
\noindent
$H_0$ : the two samples are drawn from the same underlying distribution.\\
\newline
$H_1$ : the two samples are drawn from different underlying distributions. \\

The test proceeds by pooling both samples and ranking all $n_1 + n_2$ observations
in ascending order, with tied values assigned the average of their ranks. The
Mann--Whitney U statistic for the first sample is computed as
\begin{equation}
    U_1 = R_1 - \frac{n_1(n_1+1)}{2},
    \label{eq:mannwhitney}
\end{equation}
where $R_1$ is the sum of ranks assigned to observations in the consistent-class
sample. The corresponding statistic for the erratic-class sample, $U_2$, is obtained
analogously, with $U_1 + U_2 = n_1 n_2$. The smaller of $U_1$ and $U_2$ is compared
against critical values from the Mann--Whitney distribution (or, for the sample
sizes used here, against a normal approximation with continuity correction) to
obtain a two-tailed $p$-value.

Given the large sample sizes involved (up to 29,026 locations per test), the normal
approximation to the U statistic is used throughout, with the test statistic
\begin{align}
    z &= \frac{U_1 - \mu_U}{\sigma_U}, \label{eq:zscore}\\
    \mu_U &= \frac{n_1 n_2}{2}, \\
    \sigma_U &= \sqrt{\frac{n_1 n_2 (n_1+n_2+1)}{12}}.
\end{align}
converted to a $p$-value via the standard normal cumulative distribution function.
A significance threshold of $\alpha = 0.05$ is used throughout; in practice, all
reported tests yield $p \approx 0$, well below this threshold, indicating that the
observed differences between the consistent and erratic classes are highly unlikely
to have arisen by chance under $H_0$.

We note that the Mann--Whitney U test evaluates whether one distribution is
stochastically greater than the other rather than directly comparing means or
medians. Given the large sample sizes used in this study, even small differences in
distribution can yield statistically significant results, and the effect should
therefore be interpreted alongside the visual separation shown in
\cref{fig:metrics,fig:graph_metrics} rather than the $p$-value alone.

\bibliography{aipsamp}

@misc{mosdac,
    author       = {{MOSDAC}},
    title        = {{GSMaP\_ISRO}: Global Satellite Mapping of Precipitation,
                    ISRO Product},
    howpublished = {MOSDAC Open Data Portal,
                    \url{https://www.mosdac.gov.in/opendata/GSMaP_SAC_RAIN}},
    year         = {2024},
    note         = {Accessed: 2025}
}

@article{kumar2022,
  title={Long-term high-resolution gauge adjusted satellite rainfall product over India},
  author={Kumar, Prashant and Varma, Atul K and Kubota, Takuji and Yamaji, Moeka and Tashima, Tomoko and Mega, Tomoaki and Ushio, Tomoo},
  journal={Earth and Space Science},
  volume={9},
  number={12},
  pages={e2022EA002595},
  year={2022},
  publisher={Wiley Online Library}
}

@article{kumar2024,
  title={Long-term assessment of ERA5 reanalysis rainfall for lightning events over India observed by tropical rainfall measurement mission lightning imaging sensor},
  author={Kumar, Prashant and Srivastava, Shailendra S and Jivani, Nirav and Varma, Atul K and Yokoyama, Chie and Kubota, Takuji},
  journal={Quarterly Journal of the Royal Meteorological Society},
  volume={150},
  number={761},
  pages={2472--2488},
  year={2024},
  publisher={Wiley Online Library}
}

@article{kumar2025,
  title={Long-term evaluation of CHIRPS and GSMaP\_ISRO rainfall estimates for Indian summer monsoon (2000--2022)},
  author={Kumar, Prashant and Gupta, Ujjwal K and Yamamoto, Munehisa K and Kubota, Takuji},
  journal={Remote Sensing Applications: Society and Environment},
  volume={39},
  pages={101627},
  year={2025},
  publisher={Elsevier}
}

@article{macharia2022,
  title={Validation and intercomparison of satellite-based rainfall products over Africa with TAHMO in situ rainfall observations},
  author={Macharia, Denis and Fankhauser, Katie and Selker, John S and Neff, Jason C and Thomas, Evan A},
  journal={Journal of Hydrometeorology},
  volume={23},
  number={7},
  pages={1131--1154},
  year={2022},
  publisher={American Meteorological Society}
}

@incollection{kubota2020,
  title={Global Satellite Mapping of Precipitation (GSMaP) products in the GPM era},
  author={Kubota, Takuji and Aonashi, Kazumasa and Ushio, Tomoo and Shige, Shoichi and Takayabu, Yukari N and Kachi, Misako and Arai, Yoriko and Tashima, Tomoko and Masaki, Takeshi and Kawamoto, Nozomi and others},
  booktitle={Satellite Precipitation Measurement: Volume 1},
  pages={355--373},
  year={2020},
  publisher={Springer}
}

@article{ghosh2016indian,
  title={Indian summer monsoon rainfall: implications of contrasting trends in the spatial variability of means and extremes},
  author={Ghosh, Subimal and Vittal, H and Sharma, Tarul and Karmakar, Subhankar and Kasiviswanathan, Kasiapillai S and Dhanesh, Y and Sudheer, KP and Gunthe, SS},
  journal={PloS one},
  volume={11},
  number={7},
  pages={e0158670},
  year={2016},
  publisher={Public Library of Science San Francisco, CA USA}
}

@article{kipf2016semi,
  title={Semi-supervised classification with graph convolutional networks},
  author={Kipf, Thomas N and Welling, Max},
  journal={arXiv preprint arXiv:1609.02907},
  year={2016}
}

@article{li2023graph,
  title={Graph neural network for spatiotemporal data: methods and applications},
  author={Li, Yun and Yu, Dazhou and Liu, Zhenke and Zhang, Minxing and Gong, Xiaoyun and Zhao, Liang},
  journal={arXiv preprint arXiv:2306.00012},
  year={2023}
}

@article{fey2019fast,
  title={Fast graph representation learning with PyTorch Geometric},
  author={Fey, Matthias and Lenssen, Jan Eric},
  journal={arXiv preprint arXiv:1903.02428},
  year={2019}
}

@article{xu2018powerful,
  title={How powerful are graph neural networks?},
  author={Xu, Keyulu and Hu, Weihua and Leskovec, Jure and Jegelka, Stefanie},
  journal={arXiv preprint arXiv:1810.00826},
  year={2018}
}

@article{van2008visualizing,
  title={Visualizing data using t-SNE.},
  author={Van der Maaten, Laurens and Hinton, Geoffrey},
  journal={Journal of machine learning research},
  volume={9},
  number={11},
  year={2008}
}

@inproceedings{romanova2024gnn,
  title={GNN Graph Classification for Time Series: A New Perspective on Climate Change Analysis},
  author={Romanova, Alex},
  booktitle={Proceedings of the 2024 9th International Conference on Machine Learning Technologies},
  pages={181--187},
  year={2024}
}

@article{sharma2026improving,
  title={Improving Indian summer monsoon rainfall prediction using deep learning up to two years in advance},
  author={Sharma, Devabrat and Das, Santu and Chakraborty, Deepayan and Mitra, Adway and Goswami, BN},
  journal={Quarterly Journal of the Royal Meteorological Society},
  volume={152},
  number={774},
  pages={e70023},
  year={2026},
  publisher={Wiley Online Library}
}

@article{sahastrabuddhe2023indian,
  title={Indian Summer Monsoon Rainfall in a changing climate: a review},
  author={Sahastrabuddhe, Rishi and Ghausi, Sarosh Alam and Joseph, Jisha and Ghosh, Subimal},
  journal={Journal of Water and Climate Change},
  volume={14},
  number={4},
  pages={1061--1088},
  year={2023},
  publisher={IWA Publishing}
}

@article{yadav2025recent,
  title={The recent trends in the Indian summer monsoon rainfall},
  author={Yadav, Ramesh Kumar},
  journal={Environment, Development and Sustainability},
  volume={27},
  number={6},
  pages={13565--13579},
  year={2025},
  publisher={Springer}
}

@article{verma2022regional,
  title={Regional modulating behavior of Indian summer monsoon rainfall in context of spatio-temporal variation of drought and flood events},
  author={Verma, Shruti and Bhatla, R and Shahi, NK and Mall, RK},
  journal={Atmospheric Research},
  volume={274},
  pages={106201},
  year={2022},
  publisher={Elsevier}
}

@article{hrudya2021review,
  title={A review on the Indian summer monsoon rainfall, variability and its association with ENSO and IOD},
  author={Hrudya, PH and Varikoden, Hamza and Vishnu, R},
  journal={Meteorology and Atmospheric Physics},
  volume={133},
  number={1},
  pages={1--14},
  year={2021},
  publisher={Springer}
}

@article{kapa2025variability,
  title={Variability in Indian Monsoon Onset: Delays, Advances, and Regional Disruptions},
  author={Kapa, Hemalatha and Bharghavi, Kandula and Reddy, Thotli Lokeswara},
  journal={Evolving Earth},
  pages={100103},
  year={2025},
  publisher={Elsevier}
}

@article{sahu2026climate,
  title={Climate long-distance interaction signature for spatio-temporal variability of Indian monsoon over the tropic of cancer: India},
  author={Sahu, Ramgopal T and Kumar, Kislay and Joshi, Saurabh S and Memon, Kashfina K and Kumar, Chandan},
  journal={Journal of the Indian Society of Remote Sensing},
  volume={54},
  number={1},
  pages={407--424},
  year={2026},
  publisher={Springer}
}

@article{kathayat2022protracted,
  title={Protracted Indian monsoon droughts of the past millennium and their societal impacts},
  author={Kathayat, Gayatri and Sinha, Ashish and Breitenbach, Sebastian FM and Tan, Liangcheng and Sp{\"o}tl, Christoph and Li, Hanying and Dong, Xiyu and Zhang, Haiwei and Ning, Youfeng and Allan, Robert J and others},
  journal={Proceedings of the National Academy of Sciences},
  volume={119},
  number={39},
  pages={e2207487119},
  year={2022},
  publisher={National Academy of Sciences}
}

@article{rajbanshi2021variability,
  title={The variability and teleconnections of meteorological drought in the Indian summer monsoon season: Implications for staple crop production},
  author={Rajbanshi, J and Das, S},
  journal={Journal of Hydrology},
  volume={603},
  pages={126845},
  year={2021},
  publisher={Elsevier}
}

@article{saini2022unraveling,
  title={Unraveling intricacies of monsoon attributes in homogenous Monsoon regions of India},
  author={Saini, Atul and Sahu, Netrananda and Duan, Weili and Kumar, Manish and Avtar, Ram and Mishra, Manoranjan and Kumar, Pankaj and Pandey, Rajiv and Behera, Swadhin},
  journal={Frontiers in Earth Science},
  volume={10},
  pages={794634},
  year={2022},
  publisher={Frontiers Media SA}
}

@article{thota2024spatial,
  title={Spatial variability and moisture tracks of Indian monsoon rainfall and extremes},
  author={Thota, Samba Siva Sai Prasad and Rajagopalan, Balaji},
  journal={Climate Dynamics},
  volume={62},
  number={9},
  pages={8961--8978},
  year={2024},
  publisher={Springer}
}

@book{wang2006asian,
  title={The asian monsoon},
  author={Wang, Bin},
  year={2006},
  publisher={Springer Science \& Business Media}
}

@article{duncan2013analysing,
  title={Analysing temporal trends in the Indian Summer Monsoon and its variability at a fine spatial resolution},
  author={Duncan, John MA and Dash, J and Atkinson, Peter M},
  journal={Climatic change},
  volume={117},
  number={1},
  pages={119--131},
  year={2013},
  publisher={Springer}
}

@article{mishra2022framework,
  title={A framework to incorporate spatiotemporal variability of rainfall extremes in summer monsoon declaration in India},
  author={Mishra, Vimal and Tiwari, Amar Deep and Kumar, Rohini},
  journal={Environmental Research Letters},
  volume={17},
  number={9},
  pages={094039},
  year={2022},
  publisher={IOP Publishing}
}

@article{dalai2025deciphering,
  title={Deciphering Long-Distance Climate Interactions: A Teleconnection-Based Analysis of Indian Summer Monsoon Variability},
  author={Dalai, Chitaranjan and Suram, Anil and Sahu, Ramgopal T and Kumar, Kislay and Ghongade, Manojkumar D and Joshi, Saurabh S and Rathnayake, Upaka},
  journal={Advances in Meteorology},
  volume={2025},
  number={1},
  pages={9725316},
  year={2025},
  publisher={Wiley Online Library}
}

@article{turner2012climate,
  title={Climate change and the South Asian summer monsoon},
  author={Turner, Andrew G and Annamalai, Hariharasubramanian},
  journal={Nature Climate Change},
  volume={2},
  number={8},
  pages={587--595},
  year={2012},
  publisher={Nature Publishing Group UK London}
}

@article{yadav2022monsoon,
  title={Monsoon variability in the Indian subcontinent—a review based on proxy and observational datasets},
  author={Yadav, Ankit and Mehta, Bulbul and Anoop, Ambili and Mishra, Praveen K},
  journal={Holocene climate change and environment},
  pages={369--390},
  year={2022},
  publisher={Elsevier}
}

@incollection{hegde2025spatio,
  title={Spatio-temporal variability and trends in rainfall associated with large-scale climate phenomena in Western Ghats of India},
  author={Hegde, Sahana and Gavaskar, SSM and Chengappa, SK},
  booktitle={Mitigation and adaptation strategies against climate change in natural systems},
  pages={205--220},
  year={2025},
  publisher={Springer}
}

@article{obata2024earth,
  title={Earth system model’s capability of predicting drought-induced crop failure},
  author={Obata, Atsushi and Tsujino, Hiroyuki},
  journal={Environmental Earth Sciences},
  volume={83},
  number={13},
  pages={417},
  year={2024},
  publisher={Springer}
}

@article{saipriya2026space,
  title={Space-time variability of rainfall attributes across urban agglomerations of Peninsular India and implications for sustainable growth},
  author={Saipriya, SR and Regonda, Satish Kumar and Rajagopalan, Balaji},
  journal={Urban Climate},
  volume={67},
  pages={102950},
  year={2026},
  publisher={Elsevier}
}

@article{narang2024artificial,
  title={Artificial intelligence predicts normal summer monsoon rainfall for India in 2023},
  author={Narang, Udit and Juneja, Kushal and Upadhyaya, Pankaj and Salunke, Popat and Chakraborty, Tanmoy and Behera, Swadhin Kumar and Mishra, Saroj Kanta and Suresh, Akhil Dev},
  journal={Scientific Reports},
  volume={14},
  number={1},
  pages={1495},
  year={2024},
  publisher={Nature Publishing Group UK London}
}

@article{patil2025enhancing,
  title={Enhancing Indian summer monsoon prediction: Deep learning approach for skillful long-lead forecasts of rainfall},
  author={Patil, Kalpesh R and Doi, Takeshi and Ratnam, JV and Behera, Swadhin K},
  journal={Applied Computing and Geosciences},
  volume={26},
  pages={100257},
  year={2025},
  publisher={Elsevier}
}

@inproceedings{hussain2021survey,
  title={A survey of rainfall prediction using deep learning},
  author={Hussain, Jamal and Zoremsanga, Chawngthu},
  booktitle={2021 3rd International Conference on Electrical, Control and Instrumentation Engineering (ICECIE)},
  pages={1--10},
  year={2021},
  organization={IEEE}
}

@inproceedings{saha2020cnn,
  title={Cnn-based forecasting of intraseasonal mean and active/break spells for indian summer monsoon},
  author={Saha, Moumita and Nanjundiah, Ravi S and Monteleoni, Claire},
  booktitle={Proceedings of the 10th International Conference on Climate Informatics},
  pages={15--21},
  year={2020}
}

@article{anirudh2025skillful,
  title={A skillful prediction of monsoon intraseasonal oscillation using deep learning},
  author={Anirudh, KM and Raj, Prasang and Sandeep, S and Kodamana, Hariprasad and Sabeerali, CT},
  journal={Journal of Geophysical Research: Machine Learning and Computation},
  volume={2},
  number={2},
  pages={e2024JH000504},
  year={2025},
  publisher={Wiley Online Library}
}

@article{bisht2026deep,
  title={Deep learning-driven mapping of pre-monsoon features for Indian summer monsoon precipitation forecasting},
  author={Bisht, Deepak Singh and Hazra, Vivekananda and Piyush, Durgesh Nandan},
  journal={Natural Hazards},
  volume={122},
  number={4},
  pages={142},
  year={2026},
  publisher={Springer}
}

@inproceedings{kumar2026rainfall,
  title={Rainfall Prediction Using Hybrid CNN--LSTM Model with Seasonal Trend Decomposition},
  author={Kumar, CV Satheesh and Ketan, S Hupesh Naga and others},
  booktitle={2026 IEEE International Conference on AI Engineering and Innovations (AIEI)},
  pages={1--5},
  year={2026},
  organization={IEEE}
}

@inproceedings{goyal2023conception,
  title={Conception of Indian Monsoon Prediction Methods},
  author={Goyal, Namita and Mahajan, Aparna N and Tripathi, KC},
  booktitle={International Conference on Communication and Intelligent Systems},
  pages={247--263},
  year={2023},
  organization={Springer}
}

@article{kumar2025forecasting,
  title={Forecasting of Southwest Indian Summer Monsoon Rainfall Using Artificial Neural Networks},
  author={Kumar, Aman and Singh, Samayveer and Malik, Aruna and others},
  journal={National Academy Science Letters},
  pages={1--9},
  year={2025},
  publisher={Springer}
}

@article{saha2021prediction,
  title={Prediction of the Indian summer monsoon using a stacked autoencoder and ensemble regression model},
  author={Saha, Moumita and Santara, Anirban and Mitra, Pabitra and Chakraborty, Arun and Nanjundiah, Ravi S},
  journal={International Journal of Forecasting},
  volume={37},
  number={1},
  pages={58--71},
  year={2021},
  publisher={Elsevier}
}

@article{saha2017deep,
  title={Deep learning for predicting the monsoon over the homogeneous regions of India},
  author={Saha, Moumita and Mitra, Pabitra and Nanjundiah, Ravi S},
  journal={Journal of Earth System Science},
  volume={126},
  number={4},
  pages={54},
  year={2017},
  publisher={Springer}
}

@article{dash2026performance,
  title={Performance of Global Climate Models in Predicting Indian Monsoon: A Systematic Evaluation},
  author={Dash, Yajnaseni and Gupta, Vinayak and Abraham, Ajith and Singh, Vivek and Kumar, Pradeep},
  journal={Physics and Chemistry of the Earth, Parts A/B/C},
  pages={104565},
  year={2026},
  publisher={Elsevier}
}

@article{sharma2022mechanism,
  title={Mechanism for high “potential skill” of Indian summer monsoon rainfall prediction up to two years in advance},
  author={Sharma, Devabrat and Das, Santu and Saha, Subodh K and Goswami, BN},
  journal={Quarterly Journal of the Royal Meteorological Society},
  volume={148},
  number={749},
  pages={3591--3603},
  year={2022},
  publisher={Wiley Online Library}
}

@article{kumar2022deep,
  title={Deep learning based short-range forecasting of Indian summer monsoon rainfall using earth observation and ground station datasets},
  author={Kumar, Bipin and Abhishek, Namit and Chattopadhyay, Rajib and George, Sandeep and Singh, Bhupendra Bahadur and Samanta, Arya and Patnaik, BSV and Gill, Sukhpal Singh and Nanjundiah, Ravi S and Singh, Manmeet},
  journal={Geocarto International},
  volume={37},
  number={27},
  pages={17994--18021},
  year={2022},
  publisher={Taylor \& Francis}
}

@article{bajpai2023deep,
  title={A deep and wide neural network to predict summer monsoon rainfall using time series data},
  author={Bajpai, Vikas and Bansal, Anukriti and Dash, Subrat},
  journal={Concurrency and Computation: Practice and Experience},
  volume={35},
  number={8},
  pages={e7626},
  year={2023},
  publisher={Wiley Online Library}
}

@article{saha2016predictor,
  title={Predictor discovery for early-late Indian summer monsoon using stacked autoencoder},
  author={Saha, Moumita and Mitra, Pabitra and Nanjundiah, Ravi S},
  journal={Procedia Computer Science},
  volume={80},
  pages={565--576},
  year={2016},
  publisher={Elsevier}
}

@article{dash2024integrating,
  title={Integrating empirical analysis and deep learning for accurate monsoon prediction in Kerala, India},
  author={Dash, Yajnaseni and Abraham, Ajith},
  journal={Applied Computing and Geosciences},
  volume={24},
  pages={100211},
  year={2024},
  publisher={Elsevier}
}

@article{talan2026machine,
  title={Machine learning-based rainfall prediction across temporal scales: model benchmarking and explainability analysis},
  author={Talan, Tarik},
  journal={Stochastic Environmental Research and Risk Assessment},
  volume={40},
  number={5},
  pages={100},
  year={2026},
  publisher={Springer}
}

@article{suhas2013indian,
  title={An Indian monsoon intraseasonal oscillations (MISO) index for real time monitoring and forecast verification},
  author={Suhas, E and Neena, JM and Goswami, BN},
  journal={Climate dynamics},
  volume={40},
  number={11},
  pages={2605--2616},
  year={2013},
  publisher={Springer}
}

@article{lam2023learning,
  title={Learning skillful medium-range global weather forecasting},
  author={Lam, Remi and Sanchez-Gonzalez, Alvaro and Willson, Matthew and Wirnsberger, Peter and Fortunato, Meire and Alet, Ferran and Ravuri, Suman and Ewalds, Timo and Eaton-Rosen, Zach and Hu, Weihua and others},
  journal={Science},
  volume={382},
  number={6677},
  pages={1416--1421},
  year={2023},
  publisher={American Association for the Advancement of Science}
}

@article{zhang2023st,
  title={ST-GRF: Spatiotemporal graph neural networks for rainfall forecasting},
  author={Zhang, Fang-Hao and Shao, Zhi-Gang},
  journal={Digital Signal Processing},
  volume={136},
  pages={103989},
  year={2023},
  publisher={Elsevier}
}

@article{zheng2026mesh,
  title={Mesh interpolation graph network for dynamic and spatially irregular global weather forecasting},
  author={Zheng, Zinan and Liu, Yang and Li, Jia},
  journal={Advances in Neural Information Processing Systems},
  volume={38},
  pages={25287--25311},
  year={2026}
}

@article{yousaf2025spatio,
  title={Spatio-temporal graph neural networks to improve precipitation forecasts from numerical models: U. Yousaf et al.},
  author={Yousaf, Umair and De Rango, Alessio and Furnari, Luca and D’Ambrosio, Donato and Senatore, Alfonso and Mendicino, Giuseppe},
  journal={Soft Computing},
  volume={29},
  number={9},
  pages={4481--4494},
  year={2025},
  publisher={Springer}
}

@article{yan2025convolutional,
  title={Convolutional graph neural network with novel loss strategies for daily temperature and precipitation statistical downscaling over South China},
  author={Yan, Wenjie and Liu, Shengjun and Zou, Yulin and Liu, Xinru and Wen, Diyao and Hu, Yamin and Yang, Dangfu and Xie, Jiehong and Zhao, Liang},
  journal={Advances in Atmospheric Sciences},
  volume={42},
  number={1},
  pages={232--247},
  year={2025},
  publisher={Springer}
}

@article{blasone2025graph,
  title={Graph neural networks for hourly precipitation projections at the convection permitting scale with a novel hybrid imperfect framework},
  author={Blasone, Valentina and Coppola, Erika and Sanguinetti, Guido and Arora, Viplove and Di Gioia, Serafina and Bortolussi, Luca},
  journal={Environmental Data Science},
  volume={4},
  pages={e47},
  year={2025},
  publisher={Cambridge University Press}
}

@article{xu2024dgformer,
  title={DGFormer: a physics-guided station level weather forecasting model with dynamic spatial-temporal graph neural network},
  author={Xu, Zhewen and Wei, Xiaohui and Hao, Jieyun and Han, Junze and Li, Hongliang and Liu, Changzheng and Li, Zijian and Tian, Dongyuan and Zhang, Nong},
  journal={GeoInformatica},
  volume={28},
  number={3},
  pages={499--533},
  year={2024},
  publisher={Springer}
}

@article{sun2025utility,
  title={Utility of graph neural networks in short-to medium-range weather forecasting},
  author={Sun, Xiaoni and Li, Jiming and Zhao, Zhiqiang and Jing, Guodong and Chen, Baojun and Hu, Jinrong and Wang, Fei and Zhang, Yong},
  journal={Computers, Materials, \& Continua},
  volume={84},
  number={2},
  pages={2121},
  year={2025},
  publisher={Tech Science Press}
}

@article{chen2024coupling,
  title={Coupling physical factors for precipitation forecast in China with graph neural network},
  author={Chen, Yutong and Wang, Ya and Huang, Gang and Tian, Qun},
  journal={Geophysical Research Letters},
  volume={51},
  number={2},
  pages={e2023GL106676},
  year={2024},
  publisher={Wiley Online Library}
}

@article{li2026navigating,
  title={Navigating spatio-temporal long-short heterogeneity: a dual-stream graph neural network for sparse meteorological forecasting},
  author={Li, Sheng and Li, Qian and Zhou, Zeming and Wang, Min and Zhang, Liang and Wang, Liwen},
  journal={GIScience \& Remote Sensing},
  volume={63},
  number={1},
  pages={2680356},
  year={2026},
  publisher={Taylor \& Francis}
}

@article{keisler2022forecasting,
  title={Forecasting global weather with graph neural networks},
  author={Keisler, Ryan},
  journal={arXiv preprint arXiv:2202.07575},
  year={2022}
}

@article{ma2023histgnn,
  title={HiSTGNN: Hierarchical spatio-temporal graph neural network for weather forecasting},
  author={Ma, Minbo and Xie, Peng and Teng, Fei and Wang, Bin and Ji, Shenggong and Zhang, Junbo and Li, Tianrui},
  journal={Information Sciences},
  volume={648},
  pages={119580},
  year={2023},
  publisher={Elsevier}
}

@article{blasone2026graph,
  title={Graph Neural Networks for High-Resolution Climate Projections},
  author={Blasone, Valentina and others},
  year={2026},
  publisher={Universit{\`a} degli Studi di Trieste}
}

@article{peng2023structured,
  title={A structured graph neural network for improving the numerical weather prediction of rainfall},
  author={Peng, Xuan and Li, Qian and Chen, Lei and Ning, Xiangyu and Chu, Hai and Liu, Jinqing},
  journal={Journal of Geophysical Research: Atmospheres},
  volume={128},
  number={22},
  pages={e2023JD039011},
  year={2023},
  publisher={Wiley Online Library}
}

@article{devkota2025spatio,
  title={Spatio-Temporal Weather Prediction with Graph Neural Networks},
  author={Devkota, Siddhartha and Khadka, Avinab and Pandeya, Yagya Raj},
  journal={Journal of NAST College},
  volume={1},
  number={1-2},
  pages={101--110},
  year={2025}
}

@inproceedings{coppola2025graph,
  title={Graph neural networks based climate emulator for kilometer scale hourly precipitation: a novel hybrid imperfect approach},
  author={Coppola, Erika and Blasone, Valentina and Di Gioia, Serafina and Sanguinetti, Guido and Arora, Viplove and Bortolussi, Luca},
  booktitle={EGU General Assembly Conference Abstracts},
  pages={EGU25--17645},
  year={2025}
}

@article{bhandari2024recent,
  title={Recent advances in electrical engineering: exploring graph neural networks for weather prediction in data-scarce environments},
  author={Bhandari, Harish Chandra and Pandeya, Yagya Raj and Jha, Kanhaiya and Jha, Sudan},
  journal={Environmental Research Communications},
  volume={6},
  number={10},
  pages={105010},
  year={2024},
  publisher={IOP Publishing}
}

@article{wang2026spatiotemporal,
  title={Spatiotemporal Dynamic Hypergraph Neural Network for Large-scale Long-term Online Precipitation Forecasting},
  author={Wang, Zhenghong and Huang, Wei and Lin, Zhan and Wang, Yi and Huang, Tianqiang and Zhang, Fan and Huang, Zhou and Liu, Yu},
  journal={IEEE Transactions on Geoscience and Remote Sensing},
  year={2026},
  publisher={IEEE}
}

@inproceedings{blasone2024deep,
  title={A deep learning framework to efficiently estimate precipitation at the convection permitting scale},
  author={Blasone, Valentina and Coppola, Erika and Sanguinetti, Guido and Arora, Viplove and Di Gioia, Serafina and Bortolussi, Luca},
  booktitle={ICLR 2024 Workshop on Tackling Climate Change with Machine Learning},
  year={2024}
}

@article{jones1987pictures,
  title={Pictures of relevance: A geometric analysis of similarity measures},
  author={Jones, William P and Furnas, George W},
  journal={Journal of the American society for information science},
  volume={38},
  number={6},
  pages={420--442},
  year={1987},
  publisher={Wiley Online Library}
}

@article{korenius2007principal,
  title={On principal component analysis, cosine and Euclidean measures in information retrieval},
  author={Korenius, Tuomo and Laurikkala, Jorma and Juhola, Martti},
  journal={Information Sciences},
  volume={177},
  number={22},
  pages={4893--4905},
  year={2007},
  publisher={Elsevier}
}

@article{sidorov2014soft,
  title={Soft similarity and soft cosine measure: Similarity of features in vector space model},
  author={Sidorov, Grigori and Gelbukh, Alexander and G{\'o}mez-Adorno, Helena and Pinto, David},
  journal={Computaci{\'o}n y Sistemas},
  volume={18},
  number={3},
  pages={491--504},
  year={2014},
  publisher={Instituto Polit{\'e}cnico Nacional, Centro de Investigaci{\'o}n en Computaci{\'o}n}
}

@article{chanwimalueang2017cosine,
  title={Cosine similarity entropy: Self-correlation-based complexity analysis of dynamical systems},
  author={Chanwimalueang, Theerasak and Mandic, Danilo P},
  journal={Entropy},
  volume={19},
  number={12},
  pages={652},
  year={2017},
  publisher={MDPI}
}

@article{tharwat2021classification,
  title={Classification assessment methods},
  author={Tharwat, Alaa},
  journal={Applied computing and informatics},
  volume={17},
  number={1},
  pages={168--192},
  year={2021},
  publisher={Emerald Publishing Limited}
}

@inproceedings{lipton2014optimal,
  title={Optimal thresholding of classifiers to maximize F1 measure},
  author={Lipton, Zachary C and Elkan, Charles and Naryanaswamy, Balakrishnan},
  booktitle={Joint European Conference on Machine Learning and Knowledge Discovery in Databases},
  pages={225--239},
  year={2014},
  organization={Springer}
}

@article{nachar2008mann,
  title={The Mann-Whitney U: A test for assessing whether two independent samples come from the same distribution},
  author={Nachar, Nadim and others},
  journal={Tutorials in quantitative Methods for Psychology},
  volume={4},
  number={1},
  pages={13--20},
  year={2008}
}

@article{xavier2007objective,
  title={An objective definition of the Indian summer monsoon season and a new perspective on the ENSO--monsoon relationship},
  author={Xavier, Prince K and Marzin, Charline and Goswami, Bhupendra Nath},
  journal={Quarterly Journal of the Royal Meteorological Society: A journal of the atmospheric sciences, applied meteorology and physical oceanography},
  volume={133},
  number={624},
  pages={749--764},
  year={2007},
  publisher={Wiley Online Library}
}

@article{li2022historical,
  title={A historical perspective of the La Ni{\~n}a event in 2020/2021},
  author={Li, Xiaofan and Hu, Zeng-Zhen and Tseng, Yu-heng and Liu, Yunyun and Liang, Ping},
  journal={Journal of Geophysical Research: Atmospheres},
  volume={127},
  number={7},
  pages={e2021JD035546},
  year={2022},
  publisher={Wiley Online Library}
}

@article{rajak2026comparative,
  title={Comparative analysis of long-term rainfall trends, variability and regionalization using K-means and Gaussian mixture model clustering with innovative trend analysis in rain-fed river basins},
  author={Rajak, Chandi and Das, Subhasish},
  journal={Physics and Chemistry of the Earth, Parts A/B/C},
  pages={104513},
  year={2026},
  publisher={Elsevier}
}

@article{saha2018disparity,
  title={Disparity in rainfall trend and patterns among different regions: analysis of 158 years’ time series of rainfall dataset across India: S. Saha et al.},
  author={Saha, Saurav and Chakraborty, Debasish and Paul, Ranjit Kumar and Samanta, Sandipan and Singh, SB},
  journal={Theoretical and Applied Climatology},
  volume={134},
  number={1},
  pages={381--395},
  year={2018},
  publisher={Springer}
}

@article{halder2022dynamical,
  title={Dynamical and moist thermodynamical processes associated with Western Ghats rainfall decadal variability},
  author={Halder, Subrota and Parekh, Anant and Chowdary, Jasti S and Gnanaseelan, C},
  journal={NPJ Climate and Atmospheric Science},
  volume={5},
  number={1},
  pages={8},
  year={2022},
  publisher={Nature Publishing Group UK London}
}

@article{shige2017role,
  title={Role of orography, diurnal cycle, and intraseasonal oscillation in summer monsoon rainfall over the Western Ghats and Myanmar Coast},
  author={Shige, Shoichi and Nakano, Yuki and Yamamoto, Munehisa K},
  journal={Journal of Climate},
  volume={30},
  number={23},
  pages={9365--9381},
  year={2017}
}

@article{phadtare2022froude,
  title={Froude-number-based rainfall regimes over the Western Ghats mountains of India},
  author={Phadtare, Jayesh A and Fletcher, Jennifer K and Ross, Andrew N and Turner, Andrew G and Schiemann, Reinhard KH},
  journal={Quarterly Journal of the Royal Meteorological Society},
  volume={148},
  number={748},
  pages={3388--3405},
  year={2022},
  publisher={Wiley Online Library}
}

@article{das2024dynamics,
  title={Dynamics of May ‘onset’of Indian summer monsoon over Northeast India},
  author={Das, Simanta and Goswami, Dhruba Jyoti and Mahanta, Rahul and Saha, Prolay and Goswami, BN},
  journal={Quarterly Journal of the Royal Meteorological Society},
  volume={150},
  number={764},
  pages={4533--4549},
  year={2024},
  publisher={Wiley Online Library}
}

@article{fujinami2017contrasting,
  title={Contrasting features of monsoon precipitation around the Meghalaya Plateau under westerly and easterly regimes},
  author={Fujinami, Hatsuki and Sato, Tomonori and Kanamori, Hironari and Murata, Fumie},
  journal={Journal of Geophysical Research: Atmospheres},
  volume={122},
  number={18},
  pages={9591--9610},
  year={2017},
  publisher={Wiley Online Library}
}

@article{prokop2015variation,
  title={Variation in the orographic extreme rain events over the Meghalaya Hills in northeast India in the two halves of the twentieth century: P. Prokop, A. Walanus},
  author={Prokop, Pawe{\l} and Walanus, Adam},
  journal={Theoretical and Applied Climatology},
  volume={121},
  number={1},
  pages={389--399},
  year={2015},
  publisher={Springer}
}

@article{shukla2006predictability,
  title={Predictability of seasonal climate variations: A pedagogical review},
  author={Shukla, J and Kinter, JL},
  journal={Predictability of weather and climate},
  volume={306},
  pages={341},
  year={2006},
  publisher={Cambridge University Press Cambridge, UK}
}

@article{krishnamurthy2008seasonal,
  title={Seasonal persistence and propagation of intraseasonal patterns over the Indian monsoon region},
  author={Krishnamurthy, V and Shukla, JJCD},
  journal={Climate Dynamics},
  volume={30},
  number={4},
  pages={353--369},
  year={2008},
  publisher={Springer}
}

@article{zhou2009well,
  title={How well do atmospheric general circulation models capture the leading modes of the interannual variability of the Asian--Australian monsoon?},
  author={Zhou, Tianjun and Wu, Bo and Wang, Bin},
  journal={Journal of Climate},
  volume={22},
  number={5},
  pages={1159--1173},
  year={2009}
}

\end{document}